\documentclass[journal]{IEEEtran}

\usepackage[T1]{fontenc}
\usepackage[cmex10]{amsmath}
\usepackage{color}
\usepackage{tcolorbox}
\usepackage{comment,color,soul}

\usepackage{algorithm}
\usepackage[noend]{algpseudocode}
\usepackage{pgfplots}
\makeatletter
\def\BState{\State\hskip-\ALG@thistlm}
\makeatother
\usepackage{algpseudocode}
\usepackage{cite}
\usepackage[utf8]{inputenc}
\usepackage{algorithm}
\usepackage{graphicx}
\usepackage{caption}
\usepackage{subfigure}
\usepackage{pstricks}
\usepackage{color}
\usepackage{bbm}
\usepackage{tcolorbox}
\usepackage{tikz}
\usepackage{pgfplots}
\usepackage{wrapfig}
\usepackage{lipsum}  
\usetikzlibrary{positioning}

\usepackage{amssymb}
\usepackage{amsthm}

\usepackage{relsize}
\usepackage{bbm}
\usepackage{bm}
\usepackage[font={small}]{caption}
\usepackage{comment,color,soul}
\usepackage{enumerate} 
\usetikzlibrary{arrows,automata}
\usepackage{amsfonts}

\usepackage{mathtools}
\usepackage{url}

\makeatletter
\def\thm@space@setup{\thm@preskip=2pt
        \thm@postskip=2pt \itshape}
\makeatother

\newtheoremstyle{newstyle}
{} 
{} 
{\mdseries} 
{} 
{\bfseries} 
{.} 
{ } 
{} 

\theoremstyle{newstyle}

\newtheorem*{theorem*}{Theorem}

\theoremstyle{definition}
\newtheorem*{example*}{Example}

\theoremstyle{remark}

\newtheorem*{claim*}{Claim}
\newtheorem{remark}{Remark}

\IEEEoverridecommandlockouts

\newcommand{\what}{\widehat}
\newcommand{\wtilde}{\widetilde}

\newcommand{\name}{CodedFedL }
\newcommand{\Expc}{\mathbb{E}}
\newcommand{\Prob}{\mathbb{P}}

\newcommand{\bG}{\mathbf{G}}
\newcommand{\bW}{\mathbf{W}}
\newcommand{\bw}{\mathbf{w}}
\newcommand{\bg}{\mathbf{g}}
\newcommand{\bx}{\mathbf{x}}
\newcommand{\bv}{\mathbf{v}}

\newcommand{\bl}{\boldsymbol{\ell}}
\newcommand{\by}{\mathbf{y}}
\newcommand{\bY}{\mathbf{Y}}
\newcommand{\bX}{\mathbf{X}}
\newcommand{\bom}{\boldsymbol\omega}

\newcommand{\btheta}{\boldsymbol\theta}

\newcommand{\norm}[1]{\lVert#1\rVert}

\makeatletter
\def\widebreve{\mathpalette\wide@breve}
\def\wide@breve#1#2{\sbox\z@{$#1#2$}%
     \mathop{\vbox{\m@th\ialign{##\crcr
\kern0.08em\brevefill#1{0.8\wd\z@}\crcr\noalign{\nointerlineskip}%
                    $\hss#1#2\hss$\crcr}}}\limits}
\def\brevefill#1#2{$\m@th\sbox\tw@{$#1($}%
  \hss\resizebox{#2}{\wd\tw@}{\rotatebox[origin=c]{90}{\upshape(}}\hss$}
\makeatletter

\NewDocumentCommand{\grad}{e{_^}}{%
  \mathop{}\!
  \nabla
  \IfValueT{#1}{_{\mspace{-4mu}#1}}
  \IfValueT{#2}{^{#2}}
}

\DeclareMathOperator*{\argmin}{arg\,min}

\begin{document}
 \sloppy

               \setlength{\belowcaptionskip}{-6pt}
        \setlength{\abovedisplayskip}{1mm}
        \setlength{\belowdisplayskip}{1mm}
        \setlength{\abovecaptionskip}{1mm}

        \title{Coded Computing for Low-Latency Federated Learning over Wireless Edge Networks} 
        
        \author{Saurav Prakash, Sagar Dhakal, Mustafa Akdeniz, Yair Yona, Shilpa Talwar, Salman Avestimehr, Nageen Himayat
        \thanks{Manuscript received July 17, 2020; revised September 28, 2020; accepted October 26, 2020. Digital Object Identifier 10.1109/JSAC.2020.3036961}
        \thanks{This work was part of Saurav Prakash's internship projects at Intel. A part of this work is adapted from the paper presented at the Wireless Edge Intelligence Workshop, IEEE Globecom 2019 \cite{dhakal2019codedG}. A part of this work was presented at the International Workshop on Federated Learning for User Privacy and Data Confidentiality, in Conjunction with ICML 2020 (FL-ICML'20) \cite{prakash2020coded}. We sincerely thank the editor and all the reviewers for their  valuable feedback and  detailed  comments.}
        \thanks{Saurav Prakash and Salman Avestimehr are with the Electrical Engineering Department, University of Southern California, Los Angeles, CA 90089, USA (e-mail: sauravpr@usc.edu, avestimehr@ee.usc.edu).}
        \thanks{Sagar Dhakal was with Intel Corporation and is now with Broadcom Corporation, San Jose, CA 95131, USA (e-mail: sagar.dhakal@broadcom.com).}
        \thanks{Mustafa Akdeniz, Shilpa Talwar, and Nageen Himayat are with Intel Labs, Santa Clara, CA 95054, USA (e-mail: \{mustafa.akdeniz, shilpa.talwar, nageen.himayat\}@intel.com).}
        \thanks{Yair Yona was with Intel Corporation and is now with Qualcomm, San Jose, CA 95110, USA (e-mail: yyona@qti.qualcomm.com).}
        \thanks{0733-8716 © 2020 IEEE. Personal use is permitted, but republication/redistribution requires IEEE permission. For more information, see  https://www.ieee.org/publications/rights/index.html}
       }
\maketitle

\begin{abstract}
Federated learning enables training a global model from data located at the client nodes, without data sharing and moving client data to a centralized server. Performance of federated learning in a multi-access edge computing (MEC) network suffers from slow convergence due to heterogeneity and stochastic fluctuations in compute power and communication link qualities across clients. We propose a novel \textit{coded computing} framework, CodedFedL, that injects structured coding redundancy into federated learning for mitigating stragglers and speeding up the training procedure. CodedFedL enables coded computing for non-linear federated learning by efficiently exploiting \textit{distributed kernel embedding} via random Fourier features that transforms the training task into computationally favourable distributed linear regression. Furthermore, clients generate \textit{local parity} datasets by coding over their local datasets, while the server combines them to obtain the \textit{global parity} dataset. Gradient from the global parity dataset compensates for straggling gradients during training, and thereby speeds up convergence. For minimizing the \textit{epoch deadline time} at the MEC server, we provide a tractable approach for finding the amount of coding redundancy and the number of local data points that a client processes during training, by exploiting the statistical properties of compute as well as communication delays. {We also characterize the leakage in data privacy when clients share their local parity datasets with the server. Additionally, we analyze the convergence rate and iteration complexity of CodedFedL under simplifying assumptions, by treating CodedFedL as a stochastic gradient descent algorithm. Finally, for demonstrating gains that CodedFedL can achieve in practice, we conduct numerical experiments using practical network parameters and benchmark datasets, in which CodedFedL speeds up the overall training time by up to $15\times$ in comparison to the  benchmark schemes.} \end{abstract}

\IEEEpeerreviewmaketitle

\section{Introduction}
\label{sec:introduction}
\IEEEPARstart{M}{assive} amounts of data are generated each day by the Internet of Things comprising billions of devices including autonomous vehicles, cell phones, and personal wearables \cite{saravanan2019role}. This big data has the potential to power a wide range of statistical machine learning based applications such as predicting health events like a heart attack from wearable devices \cite{din2019erratum}. To enable low-latency and efficient computing capabilities close to the user traffic, there have been significant efforts recently to develop multi-access edge computing (MEC) platforms \cite{shahzadi2017multi,kato2017priority,guo2018efficient,ai2018edge,ndikumana2019joint}. 

In classical MEC settings, client data is transferred to an underlying centralized computational infrastructure for further processing. However, client data can be of a personalized nature due to which there is an increasing privacy concern in moving the client data to a central location for any model training. For example, a person may want to use a machine learning application to predict health events like low sugar, but may not be willing to share the health records.

Federated learning framework has been recently developed to carry out machine learning tasks from data distributed at the client nodes, while the raw data is kept at the clients and never uploaded to the central server~\cite{mcmahan2016communication,konevcny2016federated}. As first formulated in \cite{mcmahan2016communication}, federated learning proceeds in two major steps. First, every client carries out a local gradient update on its local dataset. Second, a central server collects and aggregates the updates from the clients, updates the global model, and transmits it to the clients. The iterative procedure is carried out until convergence. 
\begin{figure}[htp] 
    \centering
    \includegraphics[width=.38\textwidth]{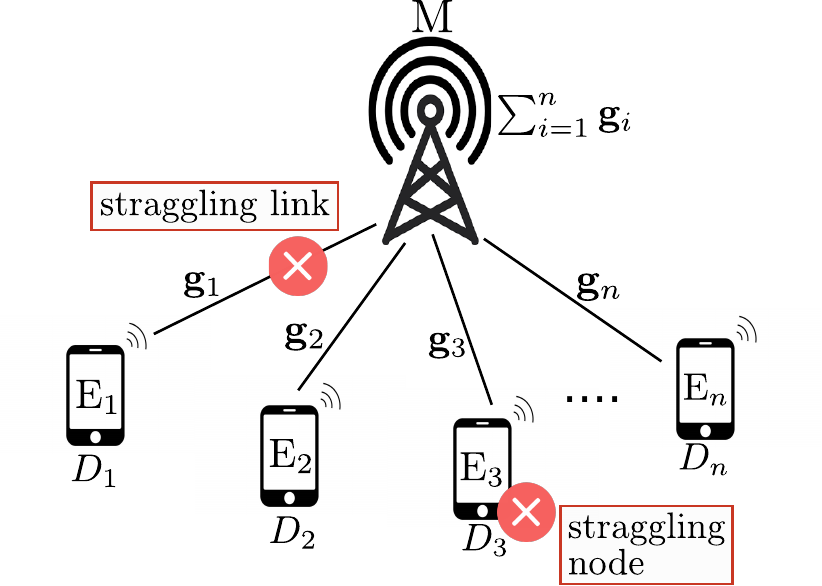}
    \caption{Illustration of the federated learning paradigm over multi-access edge computing (MEC) networks with $n$ client devices and an MEC  server. During each training round, client $\text{E}_j$ receives the latest model from server M, computes a local gradient update over its local dataset $D_j$, and communicates the gradient update to the server. Training performance is critically bottlenecked by the presence of straggling nodes and communication links.}
    \label{fig:mainfig}
\end{figure}

Implementation of federated learning in MEC networks suffers from some fundamental bottlenecks. The heterogeneity of compute and communication resources across clients makes the client selection a difficult task as the overall gradient aggregation at the MEC server can be significantly delayed by the straggling computations and communication links (see Figure \ref{fig:mainfig}). Additionally, federated learning suffers from wireless link failures during transmission. Re-transmission of messages can be done for failed communications, but it may drastically prolong the training time. Furthermore, the data distribution across clients in MEC networks is non-IID (non-independent and identically distributed), i.e. data stored locally on a device does not represent the population distribution \cite{zhao2018federated}. Thus, missing out updates from clients leads to poor convergence. 

\noindent
\textbf{CodedFedL Overview}: To overcome the aforementioned challenges, we propose \textit{CodedFedL}, a novel \textit{coded computing} framework that leverages coding theoretic ideas to inject structured redundancy in federated learning for mitigating straggling clients and communication links, and improving performance of federated learning with non-IID data. In the following, we summarize the key aspects of our proposal.
\begin{itemize}
\item \textit{Coded Computation at the MEC Server}: \name leverages the compute power of the federated learning server. Particularly, for distributed linear regression, we propose to generate masked parity data locally at each client at the start of the training procedure, by taking linear combinations of features and labels in the local dataset. The encoding coefficients are locally generated by the clients, and along with the raw client data, the encoding coefficients are not shared with the server. The local parity datasets are shared with the server, which aggregates them to obtain the composite global parity dataset.  During training, the central server obtains the coded gradient by computing the gradient over the global parity data, which compensates for the erased or delayed parameter updates from the straggling clients. The combination of coded gradient computed by the MEC server and the uncoded gradients from the non-straggling clients stochastically approximates the full gradient over the entire dataset available at the clients, thus mitigating the convergence issues arising due to missing out updates from clients when data is non-IID. 
\item \textit{Non-linear Federated Learning}: For enabling non-linear model training, we propose to have a data pre-processing step that transforms the distributed learning task into linear regression, by leveraging the popular kernel embedding based on random Fourier features (RFF)~\cite{rahimi2008random}. Each client then generates its parity dataset by taking linear combinations over its transformed features and associated labels, and the server combines them to obtain the global parity dataset. Training is then carried out with the transformed dataset at the clients and the global parity dataset at the server, as outlined in the previous bullet.
\item \textit{Optimal Load Allocation}: For obtaining the amount of coding redundancy and the number of local data points that a client processes during training, we formulate an optimization problem to find the minimum deadline time until which the MEC server should wait in each round before updating the model. We provide an analytical and tractable approach for efficiently finding the coding redundancy and load allocation that optimizes the deadline time. Our approach is based on solving a key subproblem, which can be cast as a piece-wise convex optimization problem with bounded domain and hence can be solved efficiently using standard convex optimization tools. We also derive the unique closed form solution for this subproblem for a special case in which the communication links are fully reliable with adequate error protection coding, thus covering the special case of the AWGN (Additive White Gaussian Noise) channel. 

\item \textit{Privacy Characterization}: For characterizing the privacy leakage in sharing local parity datasets with the server, we consider the case when each client utilizes an encoding matrix whose entries are independently drawn from a standard normal distribution. We consider the notion of $\epsilon$-mutual-information differential privacy (MI-DP), that is closely related to differential privacy \cite{cuff2016differential}. Specifically, we leverage the recent result in \cite{showkatbakhsh2018privacy} for the $\epsilon$-MI-DP of Gaussian random projections, and bound the leakage in a client's data privacy as a function of its database and the size of the parity dataset.

\item \textit{Convergence Analysis}: In CodedFedL, the expectation of the combination of the coded gradient and the uncoded gradients that the MEC server receives by the optimal deadline time is approximately equal to the full gradient over the entire dataset at the clients. Under simplifying assumptions, we analyze convergence and quantify the iteration complexity of CodedFedL, by treating the learning process as a stochastic gradient descent algorithm. 

\item \textit{Performance Results}: We evaluate performance gains of CodedFedL by carrying out  numerical experiments with a wireless MEC setting, benchmark datasets and non-IID data across clients. We consider the naive uncoded baseline where the server waits for all client updates, as well as the greedy uncoded baseline, where the server waits for a subset of the client updates. For achieving the same target test accuracy, CodedFedL achieves significant gains in the wall-clock training time of up to $5.8\times$ over naive uncoded, and up to $15\times$ over greedy uncoded. Furthermore, for identical number of training iterations, CodedFedL achieves almost the same test accuracy as the naive uncoded, while it outperforms greedy uncoded by an absolute accuracy margin of up to ${\sim} 13\%$, demonstrating the superiority of CodedFedL with non-IID data.
\end{itemize}

\noindent
\textbf{Related Works}
A common system approach for straggler mitigation in distributed computing has been the introduction of some form of task replication~\cite{ananthanarayanan2013effective,wang2014efficient}. Recently, coded computing strategies have been developed for injecting computation redundancy in unorthodox encoded forms to efficiently deal with communication bottleneck and system disturbances like stragglers, outages and node failures in distributed systems~\cite{li2017fundamental,lee2017speeding, reisizadeh2019coded,tandon2017gradient,karakus2017straggler,li2020coded,zhang2019improved,yan2020fundamental}. Particularly, \cite{lee2017speeding} proposed to use erasure coding for speeding up distributed matrix multiplication and linear regression tasks. Coding for heterogeneous distributed matrix multiplication is proposed in \cite{reisizadeh2019coded}, which developed an analytical method to calculate near-optimal coding redundancy. However, the entire data needs to be centrally encoded by the central server before assigning portions to compute devices. Reference \cite{tandon2017gradient} proposed a coding method over gradients for synchronous gradient descent, while \cite{karakus2019} proposed to encode over the data for avoiding the impact of stragglers for linear regression tasks. 
Many other works on coded computing for straggler mitigation in distributed learning have been proposed in the recent past \cite{ye2018communication,dhakal2019coded,yu2017polynomial,raviv2017gradient,charles2017approximate}. In all these works, the data placement and coding strategy is orchestrated by a central server. As a result, these works are not applicable in the federated learning setting, where the data is privately owned by the clients and cannot be shared with a central server. Our proposed coded computing framework, CodedFedL, provides a novel solution for leveraging coding redundancy for straggler resilient federated learning.

{Prior works that have considered one or more aspects of compute, communication and statistical heterogeneity across clients in federated learning include \cite{li2018federated,dinh2019federated,ravi2020,yoshida2020hybrid}.} In \cite{li2018federated}, a FedProx algorithm was proposed to address non-IID data across clients. However, \cite{li2018federated} did not consider variability of compute and communication capabilities across clients. Reference \cite{dinh2019federated} proposed FEDL algorithm for allocating radio resources to clients for reducing convergence time. However, we consider the MEC setting with personalized devices where the compute and communication resources of the clients cannot be tuned. In \cite{ravi2020}, important clients are selected based on compute and communication delays as well as importance of data in each round of training. In contrast, we optimize load allocation and coding redundancy only once at the start of training. Furthermore, these works do not leverage the computing capability of the MEC server. {In \cite{yoshida2020hybrid}, the authors propose a cooperative mechanism in which a fraction of clients share potentially all of their raw data with the server, which carries out gradient computations and includes them in model updates to mitigate statistical heterogeneity. However, sharing potentially all of the raw data from even a fraction of clients may not be feasible in the privacy sensitive federated learning paradigm. Additionally, the success of \cite{yoshida2020hybrid} depends on whether the clients that agree to share their raw data with the server adequately represent all the classes. This may not be practical as clients owning a certain type of data (such as users suffering from certain diseases) may not agree to participate in sharing of raw data. In CodedFedL, the server obtains a global parity dataset at the start of training via distributed encoding across client data, as each client privately encodes over its local dataset. In each training round, the server computes a coded gradient over the parity dataset that allows the central server to mitigate the impact of straggling nodes during training by stochastically approximating the gradient over the entire dataset across the clients. }

We organize the rest of our paper as follows. In Table \ref{tab:main_notations}, we list the main notations for convenience. Section \ref{sec:background} presents a technical background on federated learning, and our proposed compute and communication models for MEC. Section \ref{sec:codedfedl} describes our proposed CodedFedL scheme. In Section \ref{sec:theory}, we analyze the load allocation policy in CodedFedL. We provide the results of our numerical experiments in Section \ref{sec:experiments}, and provide our concluding remarks in Section \ref{sec:conclusions}. All technical proofs are provided in the Appendix. 

\begin{table}[htb!]
\caption{Main notations}
\label{tab:main_notations}
\centering
\begin{tabular}{|c |c|}
\hline
$n$ & number of client nodes \\ \hline
$d$ & dimension of raw feature space\\ \hline
$q$ & dimension of transformed feature space\\ \hline
$[n],\,n{\in} \mathbb{N}$ & set $\{1,\ldots,n\}$\\ \hline
$\ell_j, \,j{\in}[n]$ & number of data points in $j$-th client's dataset $D_j$\\\hline
$m$ & number of data points across all clients\\\hline
$\btheta^{r}, \,r{\in}\{1,2,\ldots\}$& global model after $r$-th training iteration\\\hline
\end{tabular}
\end{table}

\section{Problem Setup and MEC Model}
\label{sec:background}
In this section, we first describe the federated learning setting, and consider linear regression as well as non-linear regression via kernel embedding. We then present our compute and communication models for MEC.
\subsection{Federated Learning}
\label{sec:prelim}

There are $n$ client nodes, each connected to the federated learning server. Client $j{\in}[n]$ has a local dataset ${D}_j{=}({\bX}^{(j)},\bY^{(j)})$, where ${\bX}^{(j)}{\in}\mathbb{R}^{\ell_j\times d}$ and $\bY^{(j)}{\in}\mathbb{R}^{\ell_j\times c}$ denote the feature set and the label set respectively as follows:
\begin{equation}
\label{eq:localDataSet}
{\bX}^{(j)}{=}[{\bx}^{(j)T}_1,\dots,{\bx}^{(j)T}_{\ell_j}]^T,\quad     \bY^{(j)}{=}[\by^{(j)T}_1,\dots,\by^{(j)T}_{\ell_j}]^T.
\end{equation}
Here, $\ell_j{=}|{D}_j|$ denotes the number of feature-label tuples in ${D}_j$, while each data feature $\bx^{(j)}_k{\in} \mathbb{R}^{1\times d}$, and its corresponding label $\by^{(j)}_k{\in}\mathbb{R}^{1\times c}$ for $k{\in}[\ell_j]$. Client $j{\in}[n]$ does not share its dataset ${D}_j$ with the central server due to privacy concerns. 

The goal in federated learning is to train a model by leveraging the data located at the clients. Specifically, the following general problem is considered:
\begin{align}
\label{eq:mloptGeneral}
\btheta^{*}&=\argmin_{\btheta{\in} \mathcal{W}} \frac{1}{m}\sum_{j=1}^n \sum_{k=1}^{\ell_j} l\left(\btheta;(\bx^{(j)}_k,\by^{(j)}_k)\right),
\end{align}
where $l(\btheta;(\bx^{(j)}_k,\by^{(j)}_k)){\in}\mathbb{R}$ is the predictive loss associated with  $(\bx^{(j)}_k,\by^{(j)}_k)$ for model parameter $\btheta{\in} \mathcal{W}$, $m{=}\sum_{j=1}^n \ell_j$ denotes the total size of the dataset distributed across the clients, and $\mathcal{W}$ denotes the model parameter space. The solution to (\ref{eq:mloptGeneral}) is obtained via an iterative training procedure involving gradient descent. Specifically, in iteration $(r{+}1)$, server shares the current model $\btheta^{(r)}$ with the clients. Client $j{\in}[n]$ then computes the local gradient $\bg^{(j)}$ as follows:
\begin{align}
\label{eq:gradClientGeneral}
    \bg^{(j)} &= \frac{1}{\ell_j}\sum_{k=1}^{\ell_j} \grad_{\btheta}l\left(\btheta^{(r)};(\bx_k^{(j)},\by_k^{(j)})\right).
\end{align}
The server collects gradients from the clients and aggregates them to recover the gradient of the empirical loss corresponding to the entire distributed dataset across clients as follows:
\begin{align}
\label{eq:UaggregGenral}
    \bg &= \frac{1}{m}\sum_{j=1}^n \ell_j\bg^{(j)}.
\end{align}
The server then executes the model update step as follows:
\begin{equation}
\label{eq:modelupdateGeneral}
    \btheta^{(r+1)} = \btheta^{(r)} - {\mu^{(r+1)}}{\bg}
\end{equation}
where ${\mu^{(r+1)}}$ denotes the learning rate. The iterative procedure is carried out until sufficient convergence has been achieved.

For linear regression with squared error loss, the optimization problem in (\ref{eq:mloptGeneral}) is cast as follows: 
\begin{align}
\label{eq:mloptLinear}
\btheta^{*}&=\argmin_{\btheta\in \mathbb{R}^{d\times c}} \frac{1}{2m}\sum_{j=1}^n\sum_{k=1}^{\ell_j}\norm{\bx_k^{(j)}\btheta-\by_k^{(j)}}_2^2,\nonumber\\
&=\argmin_{\btheta\in \mathbb{R}^{d\times c}} \frac{1}{m}\sum_{j=1}^n \frac{1}{2}\norm{\bX^{(j)}\btheta-\bY^{(j)}}_F^2,
\end{align}
and the local gradient computation at client $j{\in}[n]$ is as follows:
\begin{align}
\label{eq:gradClientLinear}
    \bg^{(j)}=&\frac{1}{2\ell_j}\grad_{\btheta}\norm{\bX^{(j)}\btheta^{(r)} -\bY^{(j)}}_F^2\nonumber\\
=&\frac{1}{\ell_j}\bX^{(j)T}(\bX^{(j)}\btheta^{(r)}-\bY^{(j)}).
\end{align}
The gradient aggregation and model update steps in (\ref{eq:UaggregGenral}) and (\ref{eq:modelupdateGeneral}) are modified accordingly.

Linear regression has been traditionally used widely in a variety of applications including weather data analysis, market research studies and observational astronomy \cite{jobson1982multivariate,isobe1990linear,hocking2013methods}. As evident from (\ref{eq:gradClientLinear}), the gradient computations involve matrix multiplications which are computationally favourable, particularly for low powered client devices with limited compute capabilities. However, in many machine learning problems, a linear model does not perform well. 

To combine the advantages of non-linear models and low complexity gradient computations in linear regression, random Fourier feature mapping (RFFM)~\cite{rahimi2008random} based kernel regression has been widely used in practice. RFFM proposed in \cite{rahimi2008random} involves explicitly constructing finite-dimensional random features from the raw features, such that the inner product of any pair of transformed features approximates the kernel evaluation corresponding to the two raw features. Specifically, features $\bv_1{\in} \mathbb{R}^{1\times d}$ and $\bv_2{\in} \mathbb{R}^{1\times d}$ are mapped to $\what{\bv}_1$ and $\what{\bv}_2$ using a feature generating function $\phi:\mathbb{R}^{1\times d}{\rightarrow} \mathbb{R}^{1\times q}$. The RFF mapping approximates a positive definitive kernel function $K:\mathbb{R}^{1\times d}\times \mathbb{R}^{1\times d}{\rightarrow} \mathbb{R}$ as represented below:
\begin{align}
    K(\bv_1,\bv_2) \approx \what{\bv}_1\what{\bv}_2^T=\phi(\bv_1)\phi(\bv_2)^T.
\end{align}
In Section \ref{sec:kernelembed}, we propose how to carry out distributed kernel embedding, so that client $j{\in}[n]$ transforms its dataset $D_j{=}(\bX^{(j)},\bY^{(j)})$ to $\what{D}_j{=}(\what{\bX}^{(j)},\bY^{(j)})$, where $\what{\bX}^{(j)}{=}[\phi({\bx_1^{(j)}})^T,\ldots,\phi({\bx_{\ell_j}^{(j)}})^T]^T$ denotes the transformed feature set. Training is then performed via linear regression over the transformed data located at the clients, i.e., the optimization problem in  (\ref{eq:mloptGeneral}) is cast as follows:
\begin{align}
\label{eq:mloptRFF}
\btheta^{*}&=\argmin_{\btheta\in \mathbb{R}^{q\times c}} \frac{1}{2m}\sum_{j=1}^n\norm{\what{\bX}^{(j)}\btheta-\bY^{(j)}}_F^2,
\end{align}
while the gradient computation in iteration $(r{+}1)$ at client $j{\in}[n]$ is as follows:
\begin{align}
\label{eq:gradClientRFF}
\bg^{(j)}=&\frac{1}{\ell_j}\what{\bX}^{(j)T}(\what{\bX}^{(j)}\btheta^{(r)}-\bY^{(j)}).
\end{align}
Gradient aggregation and model update steps are then carried out at the server according to (\ref{eq:UaggregGenral}) and (\ref{eq:modelupdateGeneral}) respectively.
\begin{remark}
Training and inference with random features have been shown to work considerably well in practice~\cite{rahimi2008random,rahimi2008uniform,shahrampour2018data, kar2012random, rick2016random}. Hence, without loss of generality, we focus on non-linear federated learning via kernel regression with RFF mapping in the remaining part of our paper. All results easily generalize to plain linear regression.
\end{remark}
To capture the heterogeneity and stochastic nature of compute and communication capabilities across clients in MEC networks, we consider probabilistic models as described next.

\subsection{Compute and Communication Models}
\label{subsec:delayModel}
To statistically represent compute heterogeneity, we assume a shifted exponential model for local gradient computation. Specifically, the computation time for $j$-th client is given by a shifted exponential random variable $T_{cmp}^{(j)}$ as follows:
\begin{equation}
\label{eq:computeModel}
T_{cmp}^{(j)} = T_{cmp}^{(j,1)} + T_{cmp}^{(j,2)}.
\end{equation}
\noindent Here, $T_{cmp}^{(j,1)}{=}\frac{\wtilde{\ell}_j}{\mu_j}$ denotes the time in seconds to process the partial gradient over $\wtilde{\ell}_j$ data points, where data processing rate is $\mu_j$ data points per second, and $\wtilde{\ell}_j$ is bounded by the size of the local dataset $\what{D}_j$, i.e., $\wtilde{\ell}_j{\leq}\ell_j$. $T_{cmp}^{(j,2)}$ is the random variable denoting the stochastic component of compute time coming from random memory access during read/write cycles associated with Multiply-Accumulate (MAC) operations, where we assume an exponential distribution for $T_{cmp}^{(j,2)}$, i.e., $p_{T_{cmp}^{(j,2)}}(t){=}\gamma_j e^{-\gamma_j t},\,t{\geq} 0$, where $\gamma_j{=}\frac{\alpha_j\mu_j}{\wtilde{\ell}_j}$. The parameter $\alpha_j{>}0$ controls the ratio of the time spent in computing to the average time spent in memory access.

In addition to the local computation time $T_{cmp}^{(j)}$, the overall execution time for $j$-th client during  $(r+1)$-th epoch includes  $T_{com-d}^{(j)}$, time to download $\btheta^{(r)}$ from the server, and $T_{com-u}^{(j)}$, time to upload the partial gradient $\bg^{(j)}$ to the server. We assume that communications between server and clients take place over wireless links that fluctuate in quality. {It is a typical practice to model the wireless link between server and $j$-th client by a tuple $(\eta_j,p_j)$, where $\eta_j$ and $p_j$ denote the achievable data rate (in bits per second per Hz) and link erasure probability $p_j$ \cite{3gpp,courtade2011optimal,berger2008optimizing}.} Downlink and uplink communication delays are IID random variables given as follows: \footnote{For the purpose of this article, we assume the downlink and the uplink delays to be reciprocal. Generalization of our framework to asymmetric delay model is easy to address.}
\begin{align}
\label{eq:commTime}
T_{com-d}^{(j)} = N_{j}^d \tau_j,\,T_{com-u}^{(j)} = N_{j}^u \tau_j
\end{align}
Here, $\tau_j{=}\frac{b}{\eta_j W}$ is the deterministic time to upload (or download) a packet of size $b$ bits containing partial gradient $\bg^{(j)}$ (or model $\btheta^{(r)}$) and $W$ is the bandwidth in Hz assigned to the $j$-th worker device. $N_{j}^d$ and $N_{j}^u$, that denote the number of transmissions required for successful downlink and uplink communications respectively, are distributed IID according to $Geometric(p=1-p_j)$ distribution as follows:
\begin{align}
\label{eq:commModel}
\Prob(N_j^d=x)&=\Prob(N_j^u=x)\nonumber\\ 
&= p^{x-1}_j(1-p_j), \,\, x = 1,2,3,...
\end{align}
Therefore, using (\ref{eq:computeModel}) and (\ref{eq:commTime}), the total time that the $j$-th device takes to successfully receive the latest model, complete its local gradient computation, and communicate the gradient to the central server, is as follows:
\begin{align}
\label{eq:totDelay}
T_{j} = T_{com-d}^{(j)} + T_{cmp}^{(j)} + T_{com-u}^{(j)},
\end{align}
while the average delay is given as follows:
\begin{align}
\label{eq:avgDelay}
\Expc(T_{j}) = \frac{\wtilde{\ell}_j}{\mu_j} \bigg(1 +\frac{1}{\alpha_j}\bigg) + \frac{2 \tau_j}{1-p_j}.
\end{align}

The federated learning procedure can be severely impacted by slow nodes, straggling communication links, and non-IID data across clients. In the following section, we describe our proposed coded computing framework, CodedFedL, that injects structured redundancy into the federated learning procedure over MEC networks for mitigating these challenges.
\section{Proposed \name Scheme}
\label{sec:codedfedl}
\label{sec:cgd}
\begin{figure*}
    \centering
    \includegraphics[width=.95\textwidth]{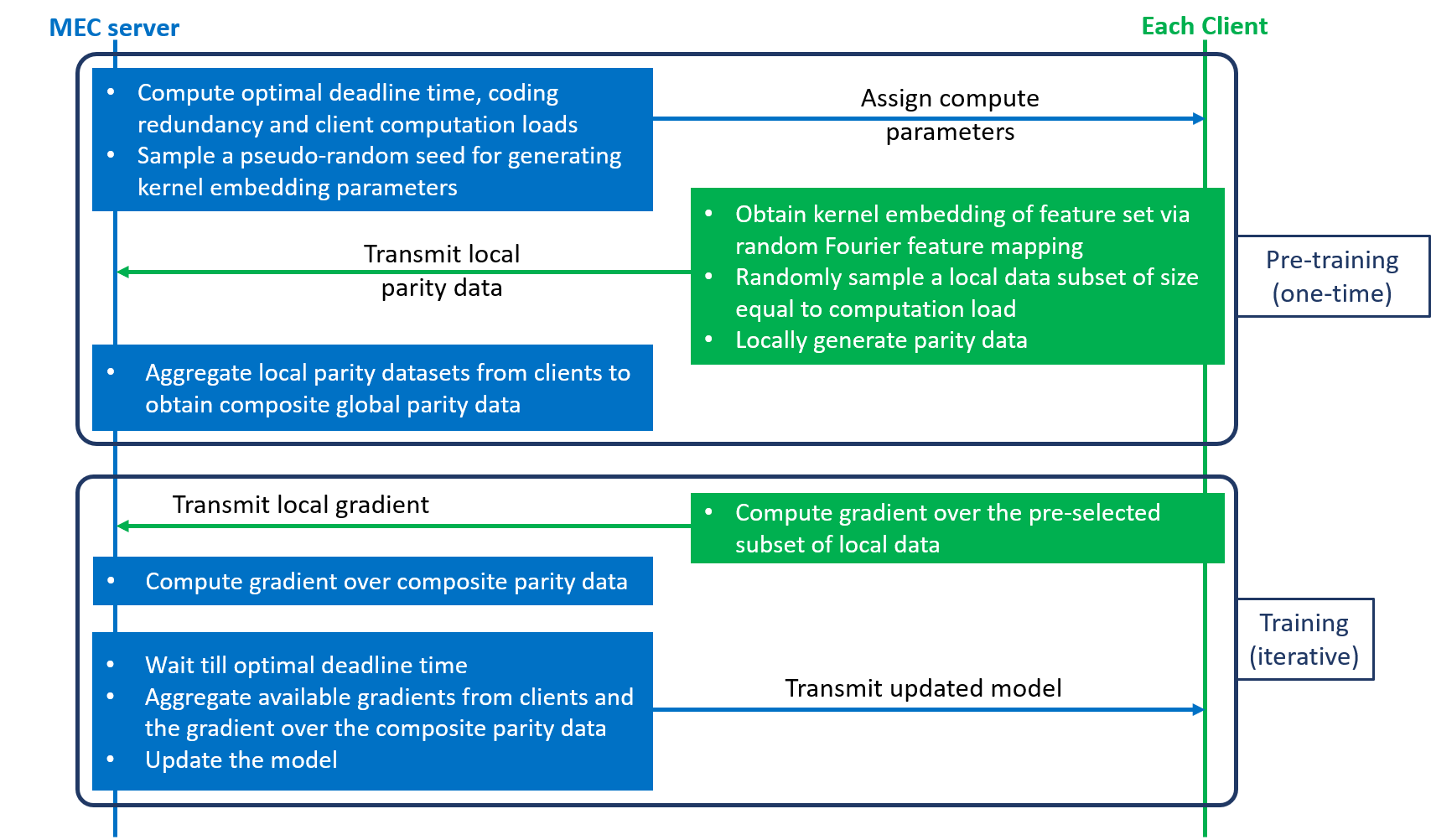}
    \caption{Overview of our proposed CodedFedL framework, illustrating the main processing steps at the MEC server and at each client.}
    \label{fig:CodedFedL_illus}
\end{figure*}

We now describe the different modules of our proposed \name scheme: distributed feature mapping for non-linear regression, distributed encoding for generating composite parity data, optimal load allocation and code design for minimizing deadline time, and modified training at the MEC server. An overview of CodedFedL is provided in Fig. \ref{fig:CodedFedL_illus}.
\subsection{Distributed Kernel Embedding}
\label{sec:kernelembed}
For combining the advantages of superior performance of non-linear models and low computational complexity of gradient computations in linear regression, we propose to leverage kernel embedding based on random Fourier feature mapping (RFFM) in federated learning. Let ${D}{=}{\cup_{j=1}^n} D_j{=}(\bX,\bY)$ denote the entire dataset located at the clients, where $\bX{\in}\mathbb{R}^{m\times d}$ and $\bY{\in}\mathbb{R}^{m\times c}$ respectively denote the combined feature and label sets at the clients as follows:
\begin{align}
    \bX=[\bX^{(1)T},\ldots,\bX^{(n)T}]^T,\quad\bY=[\bY^{(1)T},\ldots,\bY^{(n)T}]^T.
\end{align}
In this paper, we consider the commonly used kernel known as the Radial Basis Function (RBF) kernel~\cite{vert2004primer}, in which for features $\bv_1{\in}\bX$ and $\bv_2{\in}\bX$, the following relationship holds:
\begin{align}
    K(\bv_1,\bv_2) = e^{-\frac{\norm{\bv_1-\bv_2}^2}{2\sigma^2}},
\end{align}
where $\sigma$ is a kernel hyperparameter. For $i{\in}[m]$, RFFM corresponding to the RBF kernel can be carried out for feature vector $\bv_i$ as follows {(see Section V, example (a) in \cite{rahimi2008uniform})}:
\begin{align}
\label{eq:rffm}
    \what{\bv}_i&=\phi(\bv_i)\nonumber\\
    &=\sqrt{\frac{2}{q}}\left[\cos(\bv_i\bom_1+\delta_1),\ldots,  \cos(\bv_i\bom_q+\delta_q))\right]
\end{align}
where for $s{\in}[q]$, the frequency vectors $\bom_s{\in}\mathbb{R}^{d\times 1}$ are drawn independently from $\mathcal{N}(\mathbf{0}, \frac{1}{\sigma^2}\mathbf{I}_d)$, while the shift elements $\delta_s$ are drawn independently from the ${Uniform}(0,2\pi]$ distribution. Before commencing the training procedure, $j$-th client carries out RFFM on its raw feature set $\bX^{(j)}$ to obtain the transformed feature set $\what{\bX}^{(j)}{=}\phi(\bX^{(j)})$, and the training proceeds with the transformed dataset $\what{D}{=}(\what{\bX},\bY)$, where $\what{\bX}{\in} \mathbb{R}^{m\times q}$ is the matrix denoting all the transformed features across all clients. 
\begin{remark}
For distributed transformation of features at the clients, the server sends the same pseudo-random seed to every client which then obtains the samples required for RFFM in (\ref{eq:rffm}). This mitigates the need for the server to communicate the frequency vectors $\bom_1,\ldots, \bom_q$ and the shift elements $\delta_1,\ldots,\delta_q$ to each client, thus reducing the communication overhead of distributed kernel embedding significantly.
\end{remark}
Along with the computational benefits of linear regression over the transformed dataset $\what{D}{=}(\what{\bX},\bY)$, applying RFFM enables our distributed encoding strategy for creating global parity data for non-linear federated learning, as described next.  
\subsection{Distributed Encoding}
\label{encode}
To inject coding redundancy into federated learning, $j$-th client carries out random linear encoding of its transformed training dataset $\what{D}_j{=}(\what{\bX}^{(j)},{\bY}^{(j)})$. Specifically, random generator matrix $\bG_j{\in} \mathbb{R}^{u \times \ell_j}$ is used for encoding, where the row dimension $u$ denotes the \textit{coding redundancy} which is the amount of parity data to be generated at each device. Typically, $u{\ll} m$. Our strategy to find the amount $u$ of coding redundancy and the local computation loads of the clients is presented in Section \ref{sec:red}, where we describe our load allocation policy for optimizing the deadline time at the server. 

Client $j{\in}[n]$ privately draws the elements of $\bG_j$ independently from a probability distribution with mean $0$ and variance $1$. For example, it can use a standard normal $\mathcal{N}(0, 1)$ distribution, or a $Bernoulli(1/2)$ distribution with sample space $\{-1,+1\}$. Client $j$ keeps the encoding matrix $\bG_j$ private and does not share it with the server. $\bG_j$ is applied on the $weighted$ local dataset to obtain the local parity dataset ${\widebreve{D}}_j{=}(\widebreve{\bX}^{(j)},\widebreve{\bY}^{(j)})$ as follows:
\begin{align}
\label{eq:cData}
\widebreve{\bX}^{(j)} = \bG_j \bW_j \what{\bX}^{(j)},\,\,  \widebreve{\bY}^{(j)} = \bG_j \bW_j \bY^{(j)}.
\end{align}
For $\bw_j{=}[w_{j,1},\ldots,w_{j,\ell_j}]$, the weight matrix $\bW_j{=}\text{diag}(\bw_j)$ is an $\ell_j {\times} \ell_j$ diagonal matrix that weighs the training data point $(\what{\bx}^{(j)}_k,\by^{(j)}_k){\in}\what{D}_j$ with $w_{j,k}$, based on the stochastic conditions of the compute and communication resources, where $k{\in}[\ell_j]$. We defer the details of deriving $\bW_j$ to Section \ref{sec:weight}. 

The central server receives the local parity data from all client devices and combines them to obtain the {composite} global parity dataset $\widebreve{D}{=}(\widebreve{\bX},\widebreve{\bY})$, where $\widebreve{\bX} {\in} \mathbb{R}^{u \times q}$ and $\widebreve{\bY}{\in} \mathbb{R}^{u \times c}$ are the composite global parity feature set and global parity label set as follows:
\begin{align}
\label{eq:parityData}
\widebreve{\bX} = \sum_{j=1}^n \widebreve{\bX}^{(j)},\,\,  \widebreve{\bY} = \sum_{j=1}^n \widebreve{\bY}^{(j)}
\end{align}
Using (\ref{eq:cData}) and (\ref{eq:parityData}) we have the following:
\begin{align}
\label{eq:parityX}
\widebreve{\bX}= \bG \bW \what{\bX},\,
\widebreve{\bY}= \bG \bW {\bY},
\end{align}
where $\bG {=} [{\bf G}_1, \dots, {\bf G}_n]{\in} \mathbb{R}^{u\times m}$ is the global encoding matrix and ${\bf W}{\in}\mathbb{R}^{m\times m}$ is the global weight matrix given by 
$\bW{=}\text{diag}([\bw_1,\ldots,\bw_n])$. Equation (\ref{eq:parityX}) thus represents encoding over \textit{the entire decentralized dataset} $\what{D}{=}(\what{\bX}, \bY)$, performed implicitly in a distributed manner across clients. 
\begin{remark}
Although client $j{\in}[n]$ shares its locally coded dataset ${\widebreve{D}}_j{=}(\widebreve{\bX}^{(j)},\widebreve{\bY}^{(j)})$ with the central server, the local dataset $\what{D}_j$ as well as the encoding matrix $\bG_j$ are private to the client and not shared with the server. {In Appendix \ref{app:privacy}, we characterize the privacy leakage in sharing local parity dataset with the server.}  
\end{remark}
Next, we describe our load allocation policy to minimize the epoch deadline time for receiving the gradient updates from the non-straggling nodes.  
\subsection{Coding Redundancy and Load Assignment} 
\label{sec:red}
CodedFedL involves load optimization based on the statistical conditions of MEC for obtaining the minimum deadline time $t^{opt}$, and correspondingly an optimal number of data points ${\ell}^{opt}_j{\leq} \ell_j$ to be processed locally at client $j{\in}[n]$ in each round, as well as $u^{opt}{\leq} u^{max}$, the number of coded data points to be processed at the server in each round. Here, we assume that due to memory and storage constraints, the server can process a maximum of $u^{max}$ coded data points in each round. {Furthermore, for generality, we assume that the server offloads the computation to a high performance computing unit. During training, the computing unit receives the current model from the server, carries out gradient computations over its assigned coded data, and uploads the gradient to the server, where the communications to and from the server take place over a wireless channel. Therefore, to parameterize the compute and communication capabilities of the MEC server, we use similar compute and communication models as described in Section \ref{subsec:delayModel}. We let $T_{C}$ denote the random variable for the overall time spent by the computing unit in receiving the current model from the server, computing the gradient over its assigned coded data, and communicating the gradient to the server. In practice, the MEC server has dedicated, high performance and reliable cloud like compute and communication resources \cite{taleb2017multi,porambage2018survey}. Thus, in comparison to client devices in practice, MEC server has higher values for the data processing rate $\mu$ and parameter $\alpha$ in the computation model in (\ref{eq:computeModel}), a higher value of data transmission rate $\eta$ and a lower value of channel failure probability $p$ in the communication model in (\ref{eq:commModel}).}

Let ${\mathbbm{1}}_{\{T_j\leq t\}}$ be the indicator random variable denoting the event that the server receives the partial gradient over the $\wtilde{\ell}_j{\leq} {\ell}_j$ local data points from $j$-th client by the deadline time $t$, where $T_j$ denotes the total delay for client $j{\in}[n]$. To represent this contribution from $j$-th client by deadline time $t$, we use the random variable $R_j (t;\wtilde{\ell}_j){=}\wtilde{\ell}_j {\mathbbm{1}}_{\{T_j\leq t\}}$. Clearly, $R_j (t;\wtilde{\ell}_j){\in}\{0,\wtilde{\ell}_j\}$. We let $R_U(t;\wtilde{\bl}){=}\sum_{j=1}^{n} R_j(t;\wtilde{\ell}_j)$ denote the \textit{uncoded} aggregate return from the clients till deadline time $t$. Similarly, for representing the completion of the gradient computation over the parity dataset $\widebreve{D}{=}(\widebreve{\bX},\widebreve{\bY})$ within deadline time $t$, we use the random variable $R_{C}(t;u){=}u {\mathbbm{1}}_{\{T_{C}\leq t\}}$, with ${\mathbbm{1}}_{\{T_C\leq t\}}$ being the indicator random variable denoting the event that the \textit{coded gradient} is available for aggregation within deadline time $t$. Clearly, $R_C(t;u){\in}\{0,u\}$. Then, the following denotes the \textit{total aggregate return} for $t{\geq}0$:
\begin{align}
\label{eq:aggReturn}
R(t;(u,\wtilde{\mathbf{\bl}}))&=R_C(t;u)+R_U(t;\wtilde{\bl}).
\end{align}
Our goal is to optimize over $t$, $\wtilde{\bl}{=}(\wtilde{\ell}_1,\ldots,\wtilde{\ell}_n)$ and $u$ such that the optimal expected total aggregate return is $m$ for the minimum epoch deadline time, with $m$ being the total number of data points at the clients. More formally, we consider the following optimization problem:
\begin{equation}
\label{eq:prob_main}
\begin{aligned}
 &{\text{minimize}}&& t \\
&  \text{subject to} && \Expc(R(t;(u,\wtilde{\bl})))=m,\\
&&& \boldsymbol{0} \leq \wtilde{\bl}\leq (\ell_1,\ldots,\ell_n),\\
&&& 0 \leq u\leq u^{max},\\
&&& t\geq0.
\end{aligned}
\end{equation}
Let $(t^{opt},u^{opt},\bl^{opt})$ denote an optimal solution for (\ref{eq:prob_main}). In the following, we propose an efficient and tractable two-step approach for solving (\ref{eq:prob_main}).\\   
\noindent
\textbf{Step 1}: First step is to optimize $\wtilde{\bl}$ and $u$ in order to maximize the expected return $\Expc(R(t;(u,\wtilde{\bl})))$ for a fixed $t$. More precisely, for a given deadline time $t$, the goal is to solve the following for the total expected aggregate return:
\begin{equation}
\label{eq:optimLn}
\begin{aligned}
 &{\text{maximize}}&& \Expc(R(t;(u,\wtilde{\bl}))) \\
&\text{subject to} && \boldsymbol{0} \leq \wtilde{\bl}\leq (\ell_1,\ldots,\ell_n)\\
&&& 0\leq u\leq u^{max}
\end{aligned}
\end{equation}
As $\Expc(R(t;(u,\wtilde{\bl}))){=}\sum_{j=1}^{n} \Expc(R_j (t;\wtilde{\ell}_j)){+}\Expc(R_C (t;u))$, the optimization in (\ref{eq:optimLn}) can be decomposed into $(n{+}1)$ independent optimization problems, one for each client $j{\in}[n]$ as follows: 
\begin{equation}
\begin{aligned}
\label{eq:optimLj}
&{\text{maximize}}&& \Expc(R_j (t;\wtilde{\ell_j})) \\
&\text{subject to} && 0 \leq \wtilde{\ell}_j\leq \ell_j,
\end{aligned}
\end{equation}
and one for the MEC server as follows:
\begin{equation}
\begin{aligned}
\label{eq:optimLc}
&{\text{maximize}}&& \Expc(R_C (t;u)) \\
&\text{subject to} && 0 \leq u\leq u^{max},
\end{aligned}
\end{equation}

\begin{remark}
\label{rmk:concavityExpRetClnt}
In Section \ref{sec:theory}, we derive the mathematical expression for the expected return $\Expc(R_j (t;\wtilde{\ell}_j))$ for $j{\in}[n]$, and prove that it is a piece-wise concave function in $\wtilde{\ell}_j{>}0$. We also characterize the intervals within which the function is concave, and show that the boundaries are functions of the total number of transmissions needed for the successful downlink (model download) and uplink (gradient upload) communications by the deadline time $t$. Therefore, we can solve (\ref{eq:optimLj}) efficiently using any convex optimization toolbox. The analysis follows similarly for (\ref{eq:optimLc}). Therefore, (\ref{eq:optimLn}) can be solved efficiently.
\end{remark}
Let $\ell_j^*(t)$, for $j{\in}{[n]}$, and $u^*(t)$ denote optimal solutions of (\ref{eq:optimLj}) and  (\ref{eq:optimLc}) respectively, which in turn optimize (\ref{eq:optimLn}). Next, we describe the second step of our approach.
\\
\noindent
\textbf{Step 2}: Optimization of $t$ is considered in order to find the minimum deadline time $t{=}t^*$ so that the maximized expected total  aggregate return $\Expc(R(t;(u^*(t),{\bl}^{*}(t))))$ is equal to $m$. Specifically, the following optimization problem is considered:
\begin{equation}
\begin{aligned}
\label{eq:epochTime}
&{\text{minimize}}&& t \\
&\text{subject to} && \Expc(R(t;(u^*(t),{\bl}^{*}(t)))) = m,\\
&&& t \geq 0.
\end{aligned}
\end{equation} 
\begin{remark}
\label{rmk:monotoneExpRetClnt}
We show that $\Expc(R(t;(u^*(t),{\bl}^{*}(t))))$ is a monotonically increasing function in $t$ in Section \ref{sec:theory}. Therefore, (\ref{eq:epochTime}) can be efficiently solved to obtain $t^*$ using a bisection search over $t$ with a sufficiently large starting upper bound. Consequently, an optimal load allocation solution $(u^*(t^*),{\bl}^{*}(t^*))$ is obtained as a solution of (\ref{eq:optimLn}) for  $t{=}t^*$. 
\end{remark}
Our proposed two-step load allocation strategy achieves an optimal solution of (\ref{eq:prob_main}), as summarized in the following claim.
\begin{claim*}
\label{claim:optimality}
Let $(t^*,u^*(t^*),{\bl}^{*}(t^*))$ be an optimal solution obtained by solving (\ref{eq:epochTime}). Then, $t^*{=}t^{opt}$ and $(t^*,u^*(t^*),{\bl}^{*}(t^*))$ is an optimal solution of (\ref{eq:prob_main}). 
\end{claim*}
\noindent
Proof of the above claim is provided in Appendix \ref{append:proofClaim}. In the next subsection, we describe the procedure used by client $j{\in}[n]$ for obtaining the weight matrix $\bW_j$, which is used for generating the local parity dataset in (\ref{eq:cData}).
\subsection{Weight Matrix Construction} 
\label{sec:weight}
After the evaluation of the optimal load allocation $\bl^{*}(t^{*})$ for the clients described in the previous subsection, $j$-th client samples $\ell_j^{*}(t^{*})$ data points uniformly and randomly that it will process for local gradient computation in each training round. It is not revealed to the server which data points are sampled. The probability that the partial gradient computed at $j$-th client is not received at the MEC server by deadline time $t^{*}$ is $pnr_{j,1}{=}(1-\Prob(T_j{\leq}t^{*}))$. Furthermore, $(\ell_j{-}\ell_j^{*}(t^{*}))$ data points are never evaluated at the client, which implies that the probability of no return for them is $pnr_{j,2}{=}1$. 

The diagonal weight matrix $\bW_j{\in}\mathbb{R}^{\ell_j\times \ell_j}$, which is used for generating the local parity dataset in (\ref{eq:cData}), captures the absence of the updates corresponding to the different data points during the training procedure. Specifically, for $k{\in}[\ell_j]$, if data point $(\what{\bx}_k^{(j)},\by_k^{(j)})$ is among the $\ell_j^{*}(t^{*})$ data points that are to be processed at the client during gradient computation, the corresponding weight matrix coefficient is $w_{j,k}{=}\sqrt{pnr_{j,1}}$, otherwise $w_{j,k}{=}\sqrt{pnr_{j,2}}$. As we illustrate next, this weighing ensures that the combination of the coded gradient and the uncoded gradient updates from the non-straggling clients \textit{stochastically approximates} the full gradient $\bg$ in (\ref{eq:UaggregGenral}) over the entire dataset $\what{D}{=}(\what{\bX},\bY)$ distributed across the clients.

\subsection{Coded Federated Aggregation}
\label{sec:codfedagg}
{In epoch $(r{+}1)$, the MEC server sends the current model $\btheta^{(r)}$ to the clients as well as its own computing unit for gradient computations, and waits until the optimal deadline time $t^*$ before updating the model. The computing unit of the MEC server computes the coded gradient, which is the gradient over the composite parity data $\widebreve{D}{=}(\widebreve{\bX},\widebreve{\bY})$, and the MEC server weighs it with a factor of $1/{(1-pnr_C)}$, where $pnr_C{=}(1-\Prob(T_C \leq t^{*}))$ denotes the probability of no return for the coded gradient. Effectively, the coded gradient used by the MEC server during gradient aggregation can be represented as follows:
\begin{align}
\label{eq:codedGradRet}
{\bg}_C&={\mathbbm{1}}_{\{T_C\leq t^{*}\}}\frac{1}{(1-pnr_C)}\biggl(\frac{1}{u^*(t^*)}\widebreve{\bX}^{T}(\widebreve{\bX} \btheta^{(r)} - \widebreve{\bY})\biggl)\nonumber\\
&=\frac{{\mathbbm{1}}_{\{T_C\leq t^{*}\}}}{(1-pnr_C)}\what{\bf X}^T {\bf W}^T\biggl(\frac{{\bf G}^T{\bf G}}{u^*(t^*)}\biggr){\bf W}(\what{\bX}{\btheta}^{(r)}-\bY),
\end{align}
where ${\mathbbm{1}}_{\{T_C\leq t^{*}\}}$ is the indicator random variable that denotes whether the coded gradient is available for aggregation by the optimal deadline $t^*$ or not. As we describe soon, weighing the coded gradient by a factor of $\frac{1}{(1-pnr_C)}$ accounts for the averaging effect caused by the random variable ${\mathbbm{1}}_{\{T_C\leq t^{*}\}}$, and results in a stochastic approximation of the true gradient. 

Similarly, each client $j{\in}[n]$ computes its partial gradient, and the server computes a weighted combination of the uncoded gradients received from the clients by the deadline time $t^*$ as follows: 
\begin{equation}
    {\bg}_U{=}\sum_{j=1}^n \ell_j^{*}(t^{*}) {\bg}_U^{(j)},\nonumber
\end{equation}
where ${\bg}_U^{(j)}$ represents the effective gradient contribution from client $j$ by deadline time $t^*$ as follows:
\begin{equation}
\label{eq:uncodedGradClient}
{\bg}_U^{(j)}={\mathbbm{1}}_{\{T_j\leq t^{*}\}}\frac{1}{\ell_j^{*}(t^{*})}\wtilde{\bX}^{(j)T}(\wtilde{\bX}^{(j)}\btheta^{(r)}-\wtilde{\bY}^{(j)}),
\end{equation} 
Here, $\wtilde{D}_{j}{=}(\wtilde{\bX}^{(j)},\wtilde{\bY}^{(j)})$ is composed of the $\ell_j^{*}(t^{*})$ data points that $j$-th client samples for processing before training. As we show soon, no further factors similar to $\frac{1}{(1-pnr_C)}$ in (\ref{eq:codedGradRet}) are needed to be applied to the uncoded gradients as they are already accounted for during creation of the local parity datasets. Thus, the MEC server waits for the uncoded gradients from the clients and the coded gradient from its computing unit until the optimized deadline time $t^*$, and then aggregates ${\bg}_C$ and ${\bg}_U$ to obtain the \textit{coded federated gradient} as follows: 
\begin{equation}
\label{eq:cfa}
\bg_M=\frac{1}{m}(\bg_C+\bg_U).
\end{equation} 
The coded federated gradient $\bg_M$ in (\ref{eq:cfa}) stochastically approximates the full gradient ${\bg}$ in (\ref{eq:UaggregGenral}) for a sufficiently large coding redundancy $u^*(t^*)$, specifically, $\Expc(\bg_M)\approx\bg$. To verify this, we first observe that the following holds for the coded gradient ${\bg}_C$ in (\ref{eq:codedGradRet}):
\begin{align}
\label{eq:codedGrad}
\Expc({\bg}_C)&=\frac{\Expc({\mathbbm{1}}_{\{T_C\leq t^{*}\}})}{(1-pnr_C)}\what{\bf X}^T {\bf W}^T\biggl(\frac{{\bf G}^T{\bf G}}{u^*(t^*)}\biggr){\bf W}(\what{\bf X}{\btheta}^{(r)}-{\bf Y})\nonumber\\
&\overset{(a)}\approx \what{\bf X}^T {\bf W}^T{\bf W}(\what{\bf X}{\btheta}^{(r)}-{\bf Y})\nonumber\\
&= \sum_{j=1}^{n}\sum_{k=1}^{\ell_j} w_{j,k}^2 \what{\bf x}^{(j)T}_k(\what{\bf x}^{(j)}_k {\btheta}^{(r)} - {\by}^{(j)}_k).
\end{align}
In $(a)$, by using the weak law of large numbers, we have approximated the quantity $(\frac{1}{u^*(t^*)}\bG^T \bG)$ by an identity matrix. This is a reasonable approximation for a sufficiently large coding redundancy $u^*(t^*)$, since each diagonal entry in $\frac{\bG^T \bG}{u^*(t^*)}$ converges to $1$ in probability, while each non-diagonal entry converges to $0$ in probability. Furthermore, as we demonstrate via numerical experiments in Section \ref{sec:experiments}, the convergence curve as a function of iteration for CodedFedL significantly overlaps that of the naive uncoded scheme where the server waits to aggregate the results of all the clients. 

The expected aggregate gradient $\Expc(\bg_U)$ from the clients received by the server by the deadline time $t^*$ is as follows:
\begin{align}
\label{eq:receivedGrad}
&\Expc({\bg}_U)=\sum_{j=1}^n\Prob(T_j\leq t^{*}) \wtilde{\bX}^{(j)T}(\wtilde{\bX}^{(j)}\btheta^{(r)}-\wtilde{\bY}^{(j)})\nonumber\\
&\overset{(a)}=\sum_{j=1}^{n}     \sum_{\substack{k\in[\ell_j]\\(\what{\bx}^{(j)}_k,{\by}^{(j)}_k)\in\wtilde{D}_{j}}}      \Prob(T_j \leq t^*)\what{\bf x}^{(j)T}_k(\what{\bf x}^{(j)}_k {\btheta}^{(r)} - {\by}^{(j)}_k) \nonumber\\
&\overset{(b)}=\sum_{j=1}^{n}\sum_{k=1}^{\ell_j}(1-w_{j,k}^2)\what{\bf x}^{(j)T}_k(\what{\bf x}^{(j)}_k {\btheta}^{(r)} - {\by}^{(j)}_k),
\end{align}
where in $(a)$, the inner sum denotes the sum over data points in $\wtilde{D}_{j}{=}(\wtilde{\bX}^{(j)},\wtilde{\bY}^{(j)})$, while in $(b)$, all the points in the local dataset are included, with $(1-w_{j,k}^2){=}0$ for the points in the set  $\what{D}_j{\setminus}\wtilde{D}_j$. In light of (\ref{eq:codedGrad}) and (\ref{eq:receivedGrad}), it follows that $\Expc(\bg_M){\approx}{\bg}$.
\begin{remark}{In Appendix \ref{append:proofConvergence}, we perform convergence analysis of CodedFedL and find its iteration complexity under the simplifying assumption that $(\frac{1}{u^*(t^*)}\bG^T \bG){=}\mathbf{I}_m$ which based on the above analysis, implies $\Expc(\bg_M){=}{\bg}$. We bound the variance of $\bg_M$ and leverage a standard result in literature for convergence of stochastic gradient descent. The exact convergence analysis taking into account the underlying distribution of the encoding matrix will be addressed in future work.}
\end{remark}
}
In the following section, we analyze the load allocation policy of CodedFedL, demonstrating how our proposed two-step load allocation problem in (\ref{eq:epochTime}) can be solved efficiently. 
\section{Analyzing \name Load Design}
\label{sec:theory}
In this section, we demonstrate how our load allocation policy provides an efficient and tractable approach to obtaining the minimum deadline time in (\ref{eq:prob_main}). For ease of notation, we index the clients and the MEC server using $j{\in}[n+1]$ throughout this section, where $j{\in}[n]$ denotes the $n$ clients and $j{=}n{+}1$ denotes the MEC server, and use the generic term node for the MEC server as well as the clients. Likewise, $\wtilde{\ell}_{n+1}{=}u$, $\ell_{n+1}{=}u^{max}$, $\ell_{n+1}^{opt}{=}u^{opt}$, $\ell_{n+1}^{*}(t^*){=}u^{*}(t^*)$, $T_{n+1}{=}T_C$, and $R_{n+1}(t;\wtilde{\ell}_{n+1}){=}R_C(t;u)$. In the following, we present our result for the expected return $\Expc(R_j (t;\wtilde{\ell}_j))$ for node $j{\in}[n+1]$ as defined in Section \ref{sec:red}.
\begin{theorem*} 
\label{thm:expcRetWorker}
For the compute and communication models defined in (\ref{eq:computeModel}) and (\ref{eq:commModel}), let $0{\leq}\wtilde{\ell}_j{\leq}\ell_j$ be the number of data points processed by node $j{\in}[n+1]$ in each training epoch. For a deadline time of $t$ at the server, the expectation of the return $R_j(t;\wtilde{\ell}_j)=\wtilde{\ell}_j {\mathbbm{1}}_{\{T_j\leq t\}}$ satisfies the following: 
\begin{align}
    \Expc&({R_{j}(t;\wtilde{\ell}_j)})\nonumber \\
    &=\left\{
	\begin{array}{ll}
		\sum_{\nu=2}^{\nu_m}{U}\bigg(t-\frac{\wtilde{\ell}_j}{\mu_j} -\tau_j \nu\bigg)h_{\nu} f_{\nu}(t;\wtilde{\ell}_j)  & \mbox{if } \nu_m{\geq} 2   \\
		0 & \mbox{otherwise } \nonumber
	\end{array}
    \right.
\end{align}
where $U(\cdot)$ is the unit step function with $U(x){=}1$ for $x{>}0$ and $0$ otherwise,\\$f_{\nu}(t;\wtilde{\ell}_j){=}\wtilde{\ell}_j\bigg(1-e^{-\frac{\alpha_j\mu_j}{\wtilde{\ell}_j}(t-\frac{\wtilde{\ell}_j}{\mu_j}-\tau_j \nu)}\bigg)$,\\
$h_{\nu}{=}(\nu-1)(1-p_j)^2p_j^{\nu-2}$, \\
and $\nu_m{\in}\mathbb{Z}$ satisfies $t{-}\tau_j \nu_m{>}0,\, t{-}\tau_j (\nu_m+1){\leq} 0$.
\end{theorem*}
Proof of Theorem is provided in Appendix \ref{append:theoreProof}. Next, we analyze the behavior of $\Expc(R_j (t;\wtilde{\ell}_j))$ for $\nu_m{\geq}2$. For a given $t{>}0$ and $\nu{\in}\{2,\ldots,\nu_m\}$, consider $f_{\nu}(t;\wtilde{\ell}_j)$ for $\wtilde{\ell}_j{>}0$. Then, the following holds: 
\begin{align}
f''_{\nu}(t;\wtilde{\ell}_j) = -e^{-\frac{\alpha_j\mu_j}{\wtilde{\ell}_j}(t-\frac{\wtilde{\ell}_j}{\mu_j}-\nu\tau_j)}\frac{{\alpha_j^2\mu_j}^2 (t-\nu\tau_j)^2}{\wtilde{\ell}^3_j}<0.\nonumber
\end{align}
\begin{figure}[htp]
  \centering
  \subfigure[Illustration of the piece-wise concavity of the expected return $\Expc(R_j(t;\wtilde{\ell}_j))$ for $\wtilde{\ell}_j{>}0$ for node $j$.]{\includegraphics[scale=0.33]{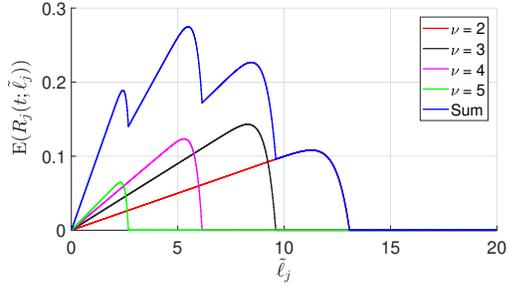}\label{fig:piecewiseConc}}\quad
  \subfigure[Illustration of the monotonic relationship between optimal expected return $\Expc(R_j (t;{\ell}^{*}_j(t)))$ and $t$ for $j$-th node.]{\includegraphics[scale=0.33]{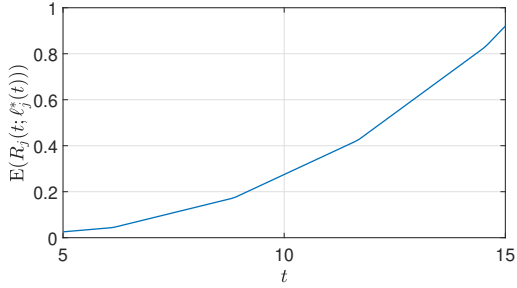}\label{fig:optretVSt}}\caption{Illustrating the properties of expected aggregate return $\Expc(R_j(t;\wtilde{\ell}_j))$ based on the result in the Theorem. We assume $p_j{=}0.9$, $\tau_j{=}\sqrt{3}$, $\mu_j{=}2$, $\alpha_j=20$, and for Fig. \ref{fig:optretVSt}, $t=10$.}\label{fig:illustrate_analysis}
\end{figure}

Thus, $f_{\nu}(t;\wtilde{\ell}_j)$ is strictly concave in the domain $\wtilde{\ell}_j{>}0$. Furthermore, $f_{\nu}(t;\wtilde{\ell}_j){\leq}0$ for $\wtilde{\ell}_j{\geq}{\mu_j(t-\tau_j \nu)}$ for $\nu{\in}\{2,\ldots,\nu_m\}$. Therefore, as highlighted in Remark \ref{rmk:concavityExpRetClnt}, the expected return $\Expc(R_j (t;\wtilde{\ell}_j))$ is piece-wise concave in $\wtilde{\ell}_j$, and the exact intervals of concavity are  $(0,\mu_j(t-\nu_m\tau_j)),\ldots,(\mu_j(t-3\tau),\mu_j(t-2\tau))$. Furthermore, $\wtilde{\ell}_j$ is upper bounded by ${\ell}_j$. Thus, for a given $t{>}0$, each of (\ref{eq:optimLj}) and (\ref{eq:optimLc}) decomposes into a finite number of convex optimization problems, that are efficiently solved in practice~\cite{boyd2004convex}. The piece-wise concave relationship between $\Expc(R_j (t;\wtilde{\ell}_j))$ and $t$ is also illustrated in Fig. \ref{fig:piecewiseConc}. 

Consider an optimal solution ${\ell}^{*}_j(t)$ for node $j{\in}[n+1]$, and the corresponding optimized expected return $\Expc(R_j(t;{\ell}^{*}_j(t)))$. Intuitively, as we increase the deadline time $t$, optimized load ${\ell}^{*}_j(t)$ should vary such that the server receives more optimal expected return from node $j$. {In Appendix \ref{append:monotonicity}, we formally prove that $\Expc(R_j(t;{\ell}^{*}_j(t)))$ is monotonically increasing in the deadline time $t$. We also illustrate this in Fig. \ref{fig:optretVSt}.} Furthermore, as the maximal expected total aggregate return $\Expc(R(t;(\ell_1^*(t),\ldots,\ell_{n+1}^*(t)))){=}\sum_{j=1}^{n+1} \Expc(R_j (t;{\ell}^{*}_j(t)))$, the maximal expected total aggregate return is also monotonically increasing in $t$. Therefore, (\ref{eq:epochTime}) can be solved efficiently using bisection search, as claimed earlier in Remark \ref{rmk:monotoneExpRetClnt}. 

When communication links do not provide time diversity, as in AWGN channel, one reliable transmission is performed at less than $10^{-5}$ bit error rate with adequate error protection coding. This motivates us to consider the special case where for each node $j{\in}[n+1]$, $p_j{=}0$, resulting in the following specialized expression for the expected return:
\begin{align}
\label{eq:awgnExpcRet}
\Expc&({R_{j}(t;\wtilde{\ell}_j)})\nonumber\\&={{U}\left(t-\frac{\wtilde{\ell}_j}{\mu_j} -2\tau_j\right)}f_{2}(t;\wtilde{\ell}_j),\nonumber\\
&={{U}\left(t-\frac{\wtilde{\ell}_j}{\mu_j} -2\tau_j\right)}    \wtilde{\ell}_j\bigg(1-e^{-\frac{\alpha_j\mu_j}{\wtilde{\ell}_j}(t-\frac{\wtilde{\ell}_j}{\mu_j}-2\tau_j )}\bigg).
\end{align}
For this special case, we have a unique closed form solution for $\bl^*(t){=}(\ell_1^*(t),\ldots,\ell_{n+1}^*(t))$, and consequently a closed form result for $\Expc(R(t;\bl^*(t)))$. Specifically, for node $j{\in}[n+1]$, we have the following one-shot solution for the optimal load $\ell_j^*(t)$:
\begin{align}
\label{eq:closedFormClientLoad2}
    \ell_j^*(t)=\left\{
	\begin{array}{ll}
		0  & \mbox{if } t\leq 2\tau_j   \\
		s_j(t-2\tau_j) & \mbox{if } 2\tau_j<t\leq \zeta_j \\
		\ell_j & \mbox{otherwise }
	\end{array}
    \right.
\end{align}
where $s_j{=}{-}\frac{\alpha_j \mu_j}{W_{-1}\left(-e^{-(1+\alpha_j)}\right)+1}$ and $\zeta_j {=} (\frac{\ell_j}{s_j}+2\tau_j)$. Here, $W_{-1}(\cdot)$ is the minor branch of the Lambert $W$-function \cite{corless1996lambertw}, which is the inverse function of $f(W){=}We^W$. Consequently, we have the following one-shot solution for the optimized return for node $j{\in}[n+1]$:
\begin{align}
\label{eq:clClRet}
    \Expc&(R_j(t;\ell_j^*(t)))\nonumber\\
    &=\left\{
	\begin{array}{ll}
		0  & \mbox{if } t\leq 2\tau_j   \\
		\wtilde{s}_j(t-2\tau_j) & \mbox{if } 2\tau_j<t\leq \zeta_j \\
		\ell_j\left(1-e^{-\frac{\alpha_j\mu_j}{\ell_j}\left(t-\frac{\ell_j}{\mu_j}-2\tau_j\right)}\right) & \mbox{otherwise }
	\end{array}
    \right.
\end{align}
where $\wtilde{s}_j{=}s_j (1-e^{-\alpha_j(\frac{\mu_j}{s_j}-1)})$. We prove (\ref{eq:closedFormClientLoad2}) and (\ref{eq:clClRet}) in Appendix \ref{append:oneshot}. Using these results, we have a closed form for the maximum expected total aggregate return from the nodes as follows:
\begin{align}
\label{eq:totalExpRet}
\Expc(R(t;\bl^*(t)))=&\sum_{\substack{j\in[n+1]\\
                  2\tau_j<t\leq \zeta_j}} \wtilde{s}_j(t-2\tau_j)\nonumber \\&+\sum_{\substack{j\in[n+1]\\               \zeta_j<t}}\ell_j\left(1-e^{-\frac{\alpha_j\mu_j}{\ell_j}\left(t-\frac{\ell_j}{\mu_j}-2\tau_j\right)}\right),
\end{align}
which is monotonically increasing in the deadline time $t$. Therefore, (\ref{eq:epochTime}) can be solved efficiently using bisection search to obtain the optimal deadline time $t^*$. 

In the next section, we present the results of our numerical experiments, which demonstrate the performance gains that CodedFedL can achieve in practice.

\section{Numerical Experiments}
\label{sec:experiments}
In this section, we demonstrate the performance gains of CodedFedL via numerical experiments. First we describe our simulation setting, and then we present the numerical results.
\subsection{Simulation Setting}
\noindent
\textbf{MEC Scenario}: We consider a wireless scenario consisting of $n{=}30$ client nodes and 1 MEC server. {For each client, the delay model described in Section \ref{subsec:delayModel} is used for the overall time in downlink (downloading the model), gradient computation, and uplink (uploading the gradient update), where the system parameters are as described next.} We use an LTE network, and assume that each client is uniformly allocated $3$ resource blocks, resulting in a maximum PHY level information rate of $216$ kbps. Note that depending on the channel conditions, the effective information rate can be lower than $216$ kbps. To model heterogeneity, we generate normalized effective information rates using $\{1, k_1, k_1^2,\ldots,k_1^{29}\}$ and assign a random permutation of them to the clients, the maximum effective information rate being $216$ kbps. Furthermore, we use the same failure probability $p_j{=}0.1$ for $j{\in}\{1,\ldots,30\}$, capturing the typical practice in wireless to adapt transmission rate for a constant failure probability \cite{3gpp}. An overhead of $10\%$ is assumed and each scalar is represented by 32 bits. The normalized processing powers are generated using $\{1, k_2, k_2^2, \ldots,k_2^{29}\}$, the maximum MAC rate being $3.072 {\times} 10^6$ MAC/s. Furthermore, we set $\alpha_j{=}2$ for $j{\in}\{1,\ldots,30\}$. We fix $(k_1,k_2){=}(0.95,0.8)$. {We assume that the MEC server has dedicated, high performance and reliable resources, so the coded gradient is available for aggregation with probability $1$ by any finite deadline time $t$, i.e. $\Prob(T_C{\leq}t){=}1$ for any $t{>}0$. Essentially, this implies $u^{opt}{=}u^{max}$. Furthermore, we let $\delta{=}u^{max}/m$ for notational convenience.}
\begin{figure*}
  \centering
  \subfigure[Test accuracy with respect to wall-clock time for naive uncoded, and CodedFedL with different $\delta$. Latency overhead due to uploading of the coded data is also highlighted.]{\includegraphics[scale=0.43]{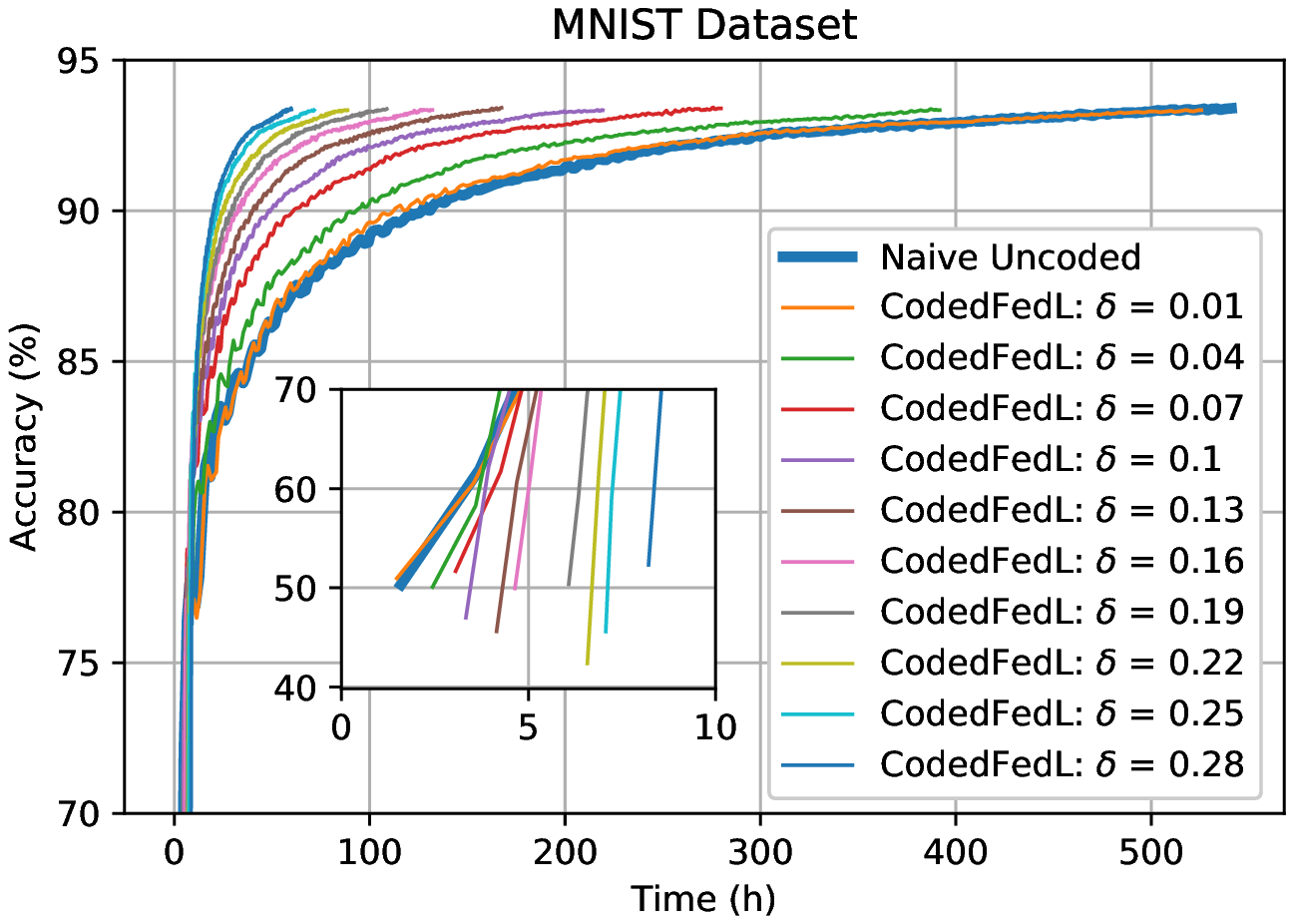}
  \label{fig:mnistCR}}\quad
  \subfigure[Test accuracy with respect to mini-batch update iteration for naive uncoded, greedy uncoded with $\psi{\in}\{0.1,0.2\}$ and CodedFedL with $\delta{\in}\{0.1,0.2\}$.]{\includegraphics[scale=0.43]{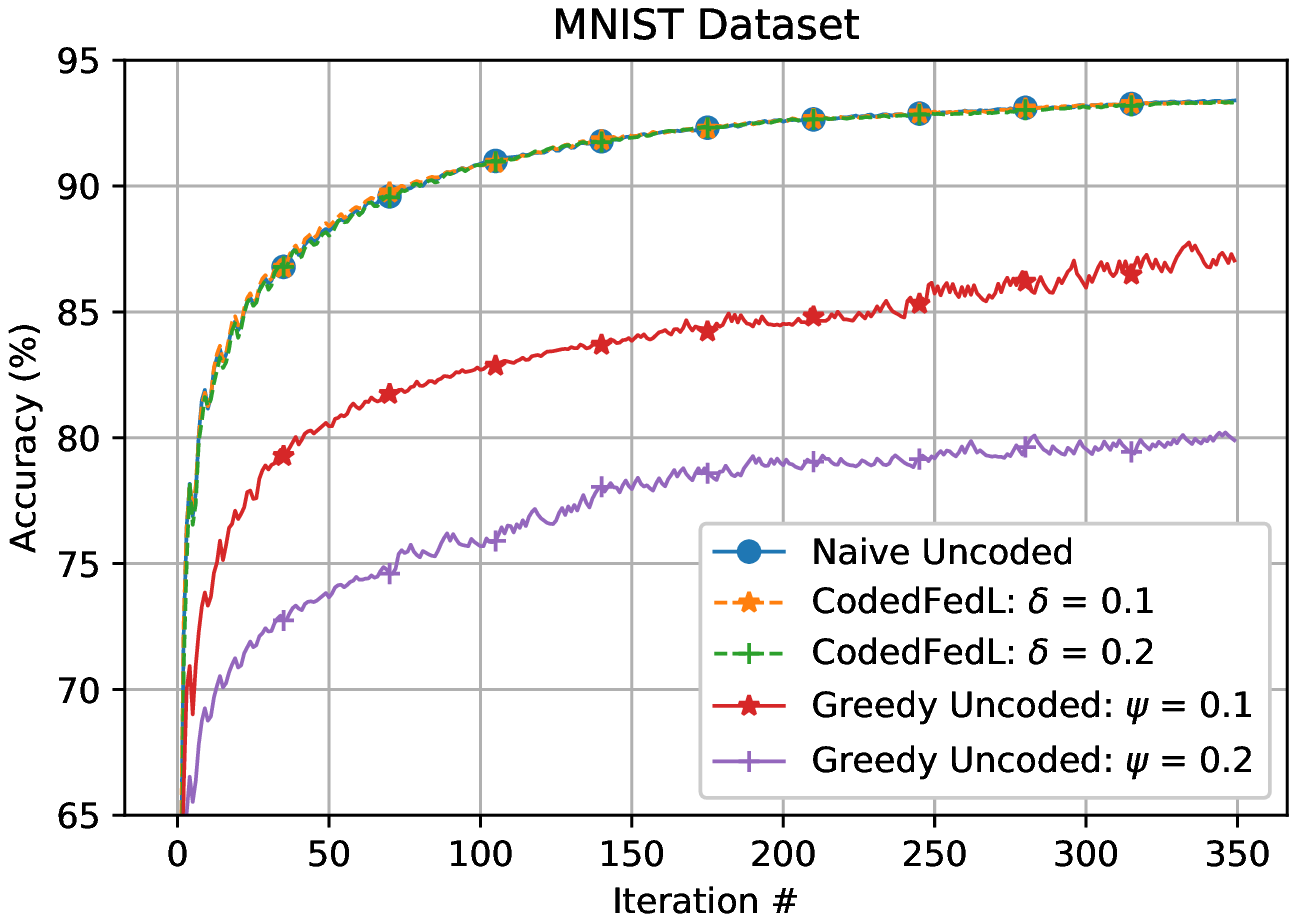}
  \label{fig:mnistWRiter}}\quad
  \subfigure[Test accuracy with respect to the wall-clock time for naive uncoded, greedy uncoded with $\psi{\in}\{0.1,0.2\}$ and CodedFedL with $\delta{\in}\{0.1,0.2\}$.]{\includegraphics[scale=0.43]{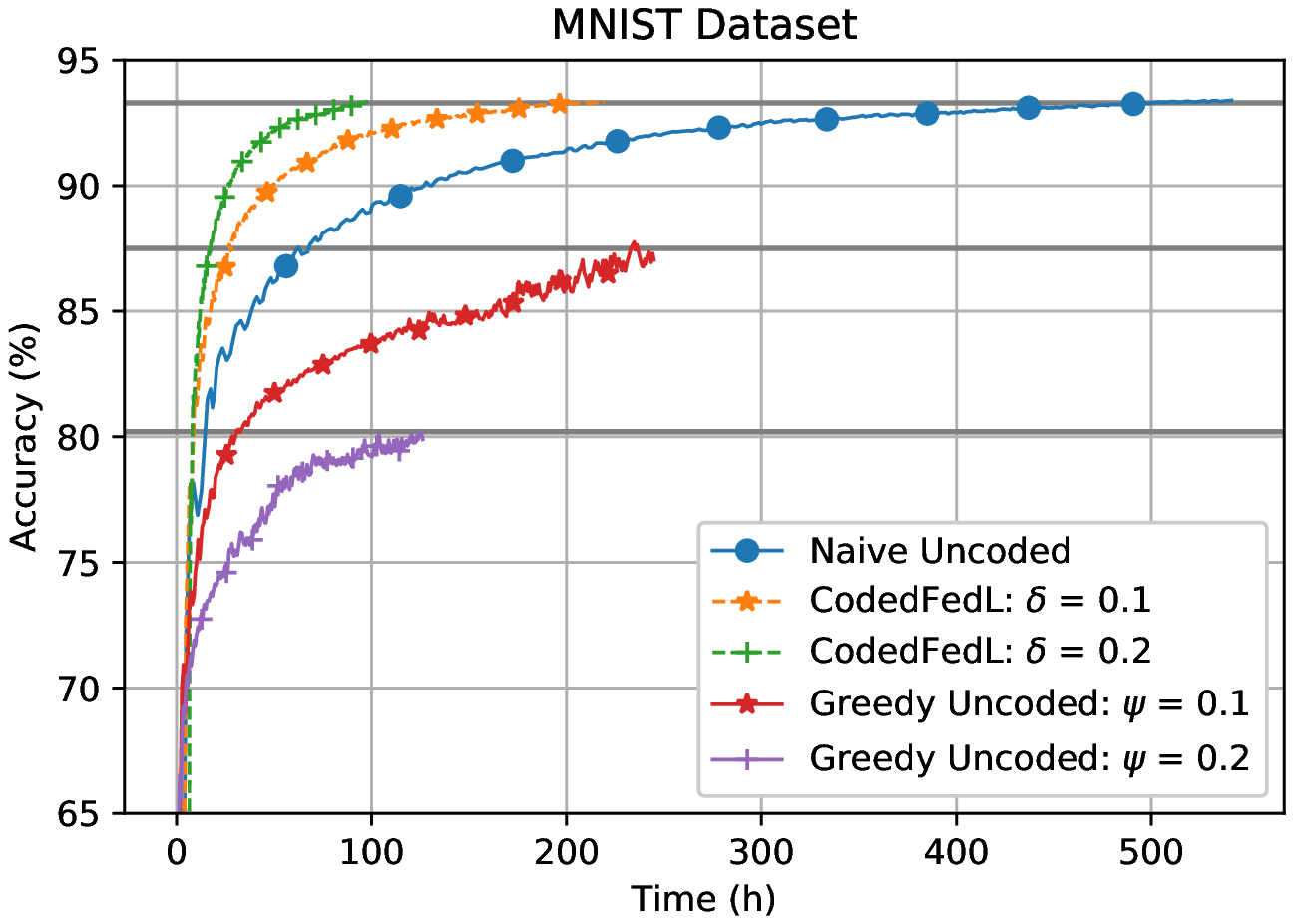}
  \label{fig:mnistWR}}
  \caption{Illustrating the results for MNIST.}
  \label{fig:results_mnist}
 \end{figure*}
 
\noindent
\textbf{Datasets and Hyperparameters}:
We consider two benchmark datasets: MNIST~\cite{lecun2010mnist} and Fashion MNIST~\cite{xiao2017fashion}. The features are vectorized, and the labels are one-hot encoded. For kernel embedding, the hyperparameters are $(\sigma,q){=}(5,2000)$. {A common practice in large-scale distributed learning is to perform training using mini-batch stochastic gradient descent (SGD), wherein the local dataset is first sorted and partitioned into mini-batches. Then, in each training iteration, each client computes gradient over a local mini-batch selected sequentially, and the model update is based on the gradient over the global mini-batch obtained by aggregating the gradients over the local mini-batches across clients. We consider the same mini-batch implementation for the uncoded schemes (as we describe in the next paragraph), while for CodedFedL, the data allocation, encoding and training modules are based on each global mini-batch.} We assign equal number of data points to each client and use a global mini-batch size of $m{=}12000$. Thus, each complete epoch over the training dataset constitutes 5 global mini-batch steps. For studying the impact of non-IID datasets across clients and demonstrate the superiority of \name in dealing with statistical heterogeneity, we first sort the training dataset according to class labels, and then partition the entire sorted training dataset into $30$ equally sized shards, each of them to be assigned to a different worker. We then sort the clients according to the expected total time using the formula in (\ref{eq:avgDelay}) with $\wtilde{\ell}_j{=}400$, i.e. the size of the local mini-batch. Then, the $30$ data shards are allocated in the order of sorted clients. For all approaches, model parameters are initialized to $0$, an initial step size of $6$ is used with a step decay of $0.8$ at epochs $40$ and $65$, while the total number of epochs is $70$. {Additionally, we use an $L_2$ regularization of $\frac{\lambda}{2}\norm{\btheta}_F^2$ with the loss defined in (\ref{eq:mloptRFF}), and the regularization parameter $\lambda$ is set to $9{\times} 10^{-6}$}. Furthermore, features are normalized to $[0,1]$ before kernel embedding. We use the RBFSampler($\cdot$) function of the sklearn library in Python for RFFM. Accuracy is reported on the test dataset for each training iteration.

\noindent{
\textbf{Schemes}:
We compare the following schemes:
\begin{itemize}
    \item Naive Uncoded: Each client computes a gradient over its local mini-batch selected sequentially, and the server waits to aggregate local gradients from all the clients.
    \item Greedy Uncoded: Clients compute gradients over their local mini-batches, and the server waits for results from the first $(1-\psi)N$ clients. This corresponds to an aggregate return of $(1-\psi)m$ from the clients. 
    \item CodedFedL: We simulate our approach described in Section \ref{sec:codedfedl}. Client $j{\in}[n]$ computes gradient over a fixed subset of $\ell_j^*(t^*)$ data points in the local mini-batch, and the server only waits till the deadline time $t^*$ before carrying out the coded federated aggregation in (\ref{eq:cfa}) corresponding to the global mini-batch for that iteration. We also include the overhead time for uploading the local parity datasets from the clients to the server.
\end{itemize}}

\begin{figure*}
  \centering
  \subfigure[Test accuracy with respect to wall-clock time for naive uncoded, and CodedFedL with different $\delta$. Latency overhead due to uploading of the coded data is also highlighted.]{\includegraphics[scale=0.43]{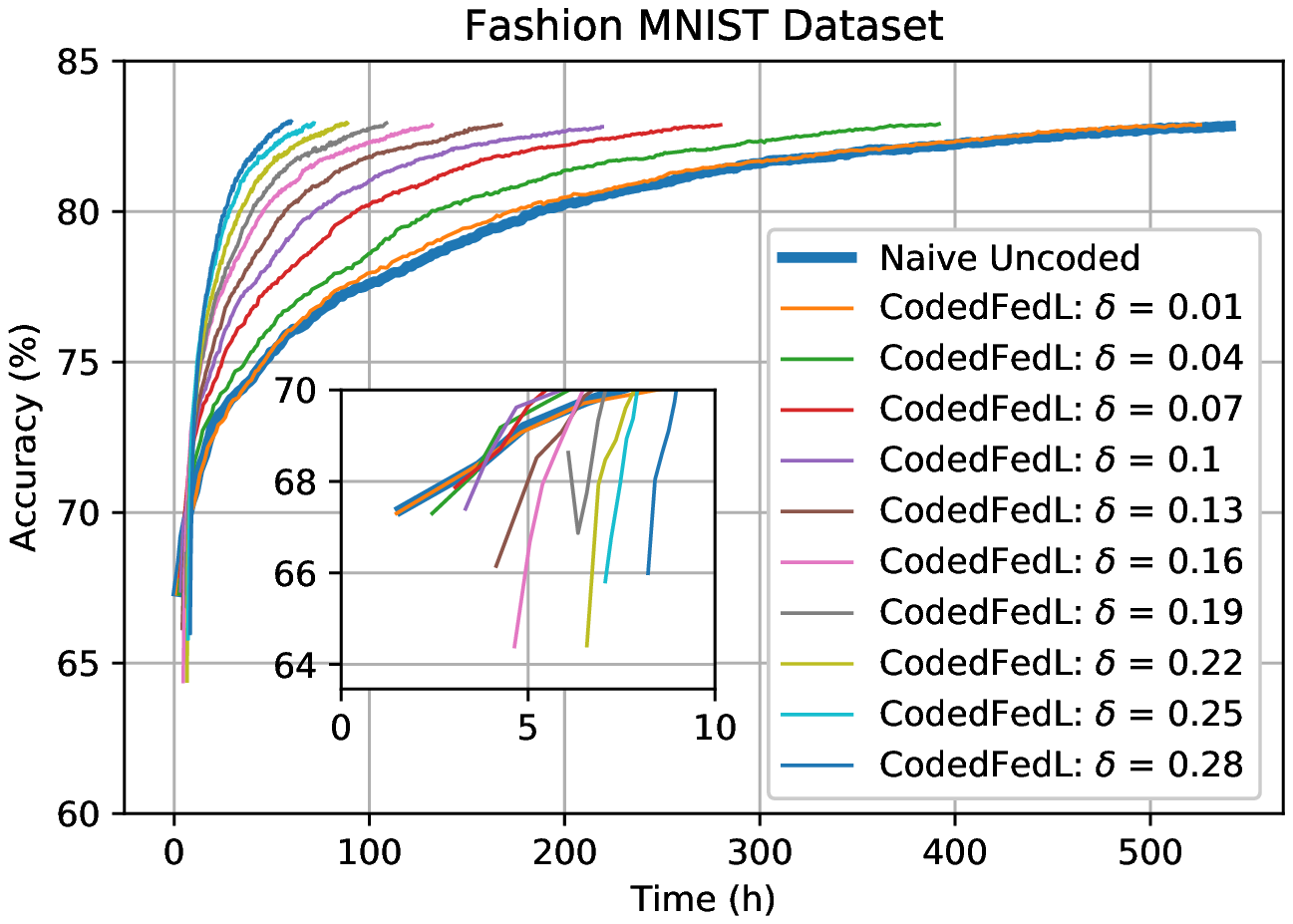}
  \label{fig:fashionCR}}\quad
  \subfigure[Test accuracy with respect to mini-batch update iteration for naive uncoded, greedy uncoded with $\psi{\in}\{0.1,0.2\}$ and CodedFedL with $\delta{\in}\{0.1,0.2\}$.]{\includegraphics[scale=0.43]{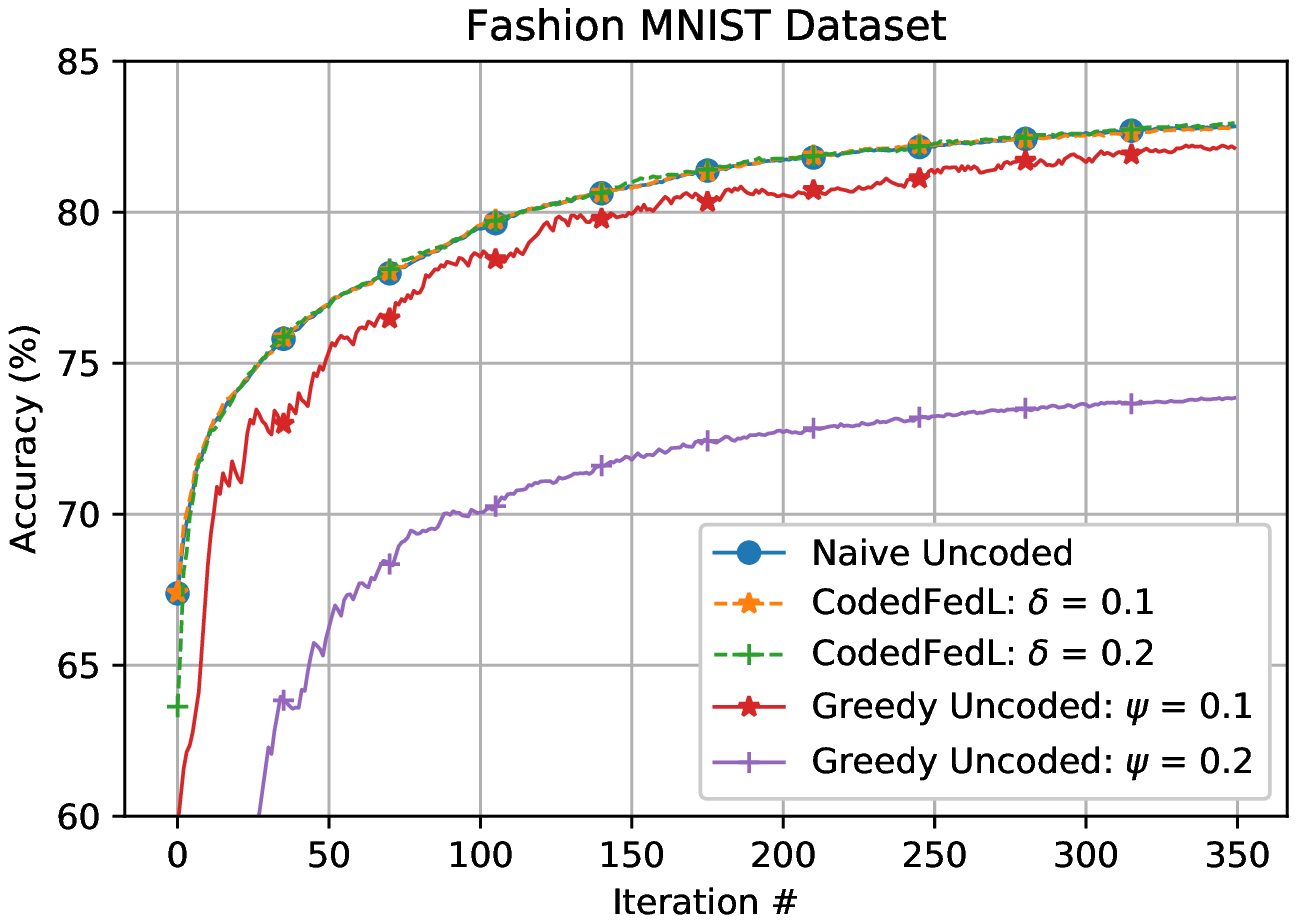}
  \label{fig:fashionWRiter}}\quad
  \subfigure[Test accuracy with respect to the wall-clock time for naive uncoded, greedy uncoded with $\psi{\in}\{0.1,0.2\}$ and CodedFedL with $\delta{\in}\{0.1,0.2\}$.]{\includegraphics[scale=0.43]{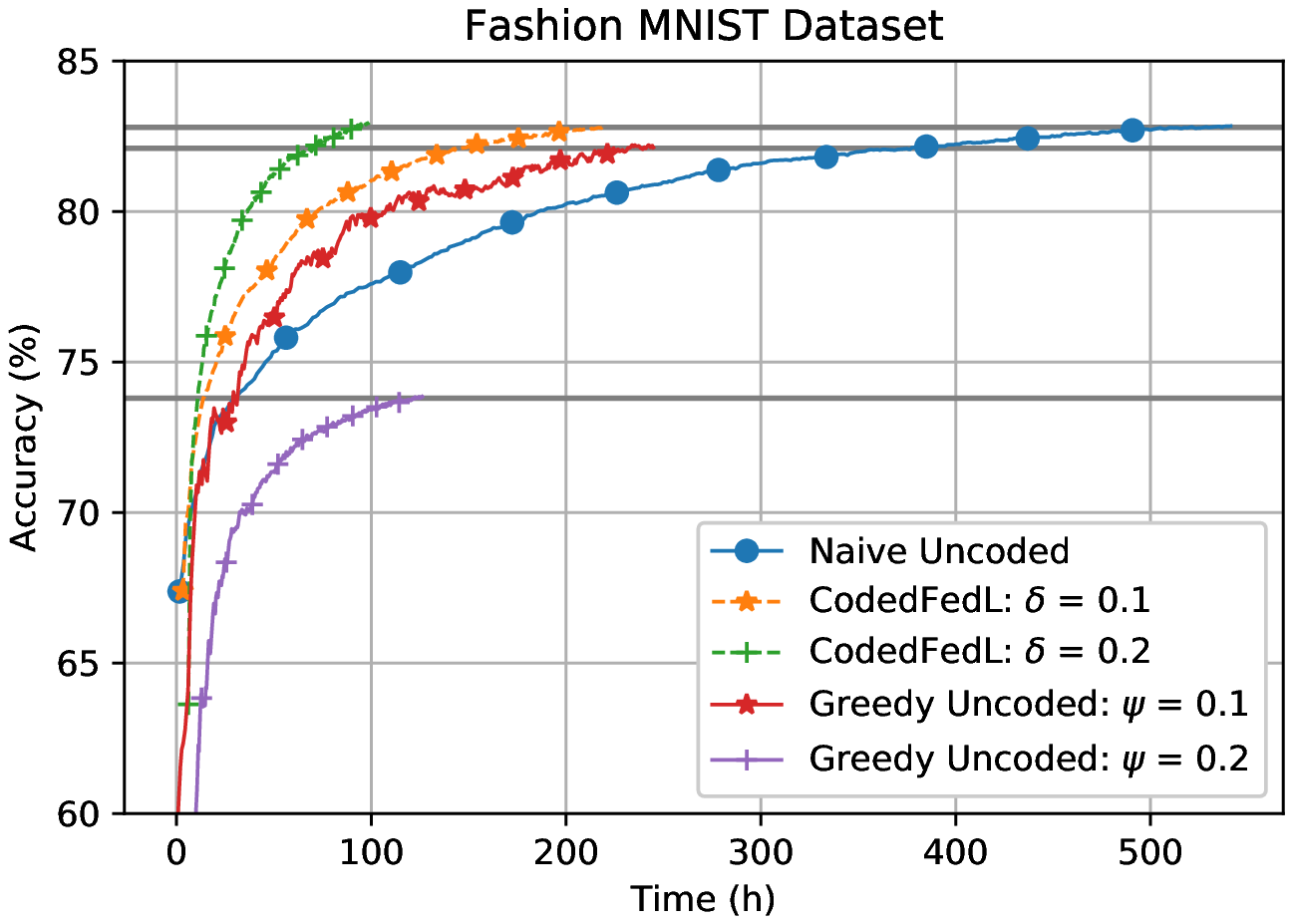}
  \label{fig:fashionWR}}
  \caption{Illustrating the results for Fashion MNIST.}
  \label{fig:results_fashion}
 \end{figure*} 
 
\subsection{Results}
Fig.~\ref{fig:results_mnist} illustrates the results for MNIST, while Fig. \ref{fig:results_fashion} illustrates the results for Fashion MNIST. In Fig. \ref{fig:mnistCR} and Fig. \ref{fig:fashionCR}, we present the generalization accuracy as a function of wall-clock time for naive uncoded and CodedFedL with different coding redundancy. Clearly, as the coding redundancy is increased by increasing $\delta$, the overall training time reduces significantly. Additionally, as highlighted by the inner figures in Fig. \ref{fig:mnistCR} and Fig. \ref{fig:fashionCR}, the initial time spent in uploading the coded data to the server generally increases with increased coding redundancy. However, the gain in training time accumulates across training iterations and the impact of this overhead becomes negligible. Furthermore, Fig.  \ref{fig:mnistCR} and Fig. \ref{fig:fashionCR} illustrate that for the same number of training iterations, CodedFedL achieves a similar generalization accuracy as the naive uncoded scheme even over a large range of coding redundancy. These plots complement our proof of the stochastic approximation of the naive uncoded gradient aggregation by the coded federated gradient aggregation in Section \ref{sec:codfedagg}, and show that for even a large coding redundancy, accuracy does not drop significantly.{ \footnote{{Using the generic local minimizer function \textsf{fminbnd} in MATLAB for solving the concave maximization subproblems, it takes lesser than $2$ minutes for implementing our two-step approach for obtaining optimal load allocation and deadline time for all values of coding redundancy considered in the simulations. Although it is negligible compared to the overall training time (see Fig. \ref{fig:mnistWR} for example), it can be further improved through an implementation specialized for convex programming.}}}

To highlight the superior performance of CodedFedL when data is non-IID, we compare the convergence plots of generalization accuracy vs training iteration for CodedFedL with $\delta{\in}\{0.1,0.2\}$ and greedy uncoded with $\psi{\in}\{0.1,0.2\}$. By design of the simulation setting, $\psi{=}0.1$ implies that for greedy uncoded, the server misses all the updates associated with a particular class in most iterations, and similarly, $\psi{=}0.2$ implies that the server misses all the updates associated with two classes in most iterations. As shown in Fig. \ref{fig:mnistWRiter} and Fig. \ref{fig:fashionWRiter}, this results in a poorer generalization performance with respect to training iteration for greedy uncoded in comparison to CodedFedL. Additionally, due to optimal load allocation, CodedFedL performs significantly better than greedy uncoded in the overall training time for identical number of training iterations, as shown in Fig.  \ref{fig:mnistWR} and Fig. \ref{fig:fashionWR}. 

Clearly, CodedFedL has significantly better convergence time than the naive uncoded and greedy uncoded approaches, and as highlighted in Section \ref{sec:codfedagg}, the coded federated gradient aggregation approximates the naive uncoded gradient aggregation well for large datasets. For further insight, let $\gamma$ be the target accuracy for a dataset, while $t_\gamma^U$, $t_\gamma^G$ and $t_\gamma^C$ respectively be the first time instants to reach the $\gamma$ accuracy for naive uncoded, greedy uncoded and CodedFedL. In Table \ref{tab:summary}, we summarize the results where $\delta{=}\psi{=}0.1$. Gains in the overall training time for \name are up to $2.5\times$ and $8.8\times$ over naive uncoded and greedy uncoded respectively. In Table \ref{tab:summary2}, we compare results for $\delta{=}\psi{=}0.2$, where the gains in the training time for the target accuracy are up to $5.4\times$ and $15\times$ over naive uncoded and greedy uncoded respectively. Also, greedy uncoded never reaches the target accuracy of $93.3\%$ for MNIST and $82.8\%$ for Fashion MNIST, hence the corresponding fields in Tables \ref{tab:summary} and \ref{tab:summary2} are empty.
\begin{table}[htb!]
\caption{Summary of Results for $\delta{=}\psi{=}0.1$}\label{tab:summary}
\centering
\begin{tabular}{l |c c c c c c}
Dataset & $\gamma$ (\%) & $t_\gamma^U$ (h) & $t_\gamma^G$ (h) & $t_\gamma^{C}$ (h) & $t_\gamma^U/t_\gamma^C$ & $t_\gamma^G/t_\gamma^C$ \\ \hline
MNIST & $93.3$ & $501$ & --- & $198$ & $2.5\times$ &--- \\ 
& $87.5$ & $63$ & $233$ & $27$ & $2.3\times$ & $8.8\times$\\ \hline
Fashion & $82.8$ & $521$ & --- & $219$ & $2.4\times$ &---\\ 
MNIST& $82.1$ & $377$ & $224$ & $145$ & $2.6\times$ & $1.6\times$\\ \hline
\end{tabular}
\end{table}

\begin{table}[htb!]
\caption{Summary of Results for $\delta{=}\psi{=}0.2$}\label{tab:summary2}
\centering
\begin{tabular}{l |c c c c c c}
Dataset & $\gamma$ (\%) & $t_\gamma^U$ (h) & $t_\gamma^G$ (h) & $t_\gamma^{C}$ (h) & $t_\gamma^U/t_\gamma^C$ & $t_\gamma^G/t_\gamma^C$ \\ \hline
MNIST & $93.3$ & $501$ & --- & $93.2$ & $5.4\times$ &--- \\ 
& $80.2$ & $15.8$ & $125$ & $8.17$ & $1.9\times$ & $15\times$\\ \hline
Fashion & $82.8$ & $521$ & --- & $90.4$ & $5.8\times$ &---\\ 
MNIST& $73.8$ & $30.6$ & $123$ & $11.1$ & $2.7\times$ & $11\times$\\ \hline
\end{tabular}
\end{table}

\section{Conclusion}
\label{sec:conclusions}
We propose CodedFedL, which is the first coding theoretic framework for straggler resilient federated learning in multi-access edge computing networks with general non-linear regression and classification tasks and non-IID data across clients. As a key component, we propose distributed kernel embedding of raw features at clients using a common pseudo-random seed across clients so that they can obtain the kernel features without having to collaborate with each other. In addition to transforming the non-linear federated learning problem into  computationally favourable distributed linear regression, kernel embedding enables our novel distributed encoding strategy that generates global parity data for straggler mitigation. The parity data allows the central server to perform gradient computations that substitute or replace missing gradient updates from straggling client devices, thus clipping the tail behavior of gradient aggregation and significantly improving the convergence performance when data is non-IID. Furthermore, there is no decoding of partial gradients required at the central server. We provide an analytical solution for load allocation and coding redundancy computation for obtaining the optimal deadline time, by utilizing statistical knowledge of compute and communication delays of the MEC nodes. Additionally, we provide privacy analysis of generating local parity datasets, and analyze convergence performance of CodedFedL under simplifying assumptions. Finally, we provide results from numerical experiments over benchmark datasets and practical network parameters that demonstrate gains of up to $15\times$ in the wall-clock  training time over benchmark schemes. 

{CodedFedL opens up many interesting future directions. As the global parity dataset is obtained by the MEC server by aggregating the local parity datasets from the clients, the encoded data of each client can be further anonymized by using secure aggregation \cite{45808}, so that the server gets to know only the global parity dataset, without knowing any individual local parity dataset. With respect to any given client $j{\in}[n]$, the server will thus receive the sum of the local parity dataset from client $j$ and a noise term, where the noise term will be the sum of the local parity datasets of the remaining clients. Exploring and characterizing this aspect of privacy is left for future study. Furthermore, the problem of characterizing the complete impact of the encoding matrix on convergence and optimizing the deadline time based on convergence criteria will be addressed in future work. Additionally, formulating and studying the load optimization problem based on outage probability for aggregate return is an interesting future work. Adapting CodedFedL to scenarios when the datasets at the clients are changing over time is a motivating future direction as well. Moreover, establishing theoretical foundations of combining coding with random Fourier feature mapping is of significant interest. Another important extension of \name is to develop coded computing solutions for federated learning for neural network workloads.}

\bibliographystyle{ieeetr}
\bibliography{biblio}
\appendices
\section{Proof of Optimality of the Two-step Approach}
\label{append:proofClaim}
Let $(t^*,u^*(t^*),{\bl}^{*}(t^*))$ be an optimal solution of (\ref{eq:epochTime}). Then, $t^*{\geq} 0$, $0{\leq} u^*(t^*){\leq} u^{max}$, $\boldsymbol{0} {\leq} {\bl}^{*}(t^*){\leq} (\ell_1,\ldots,\ell_n)$, and the following holds:
\begin{align}
\label{eq:1}
\Expc(R(t^*;(u^*(t^*),{\bl}^{*}(t^*))))&{=}m\nonumber\\
&{=}\Expc(R(t^{opt};(u^{opt},{\bl}^{opt}))).
\end{align}
Thus, $(t^*,u^*(t^*),{\bl}^{*}(t^*))$ is a feasible solution of the optimization problem in  (\ref{eq:prob_main}). Therefore, we only need to show that $t^*{=}t^{opt}$. As optimization problem in (\ref{eq:prob_main}) has a larger solution space than the two-step optimization problem in (\ref{eq:epochTime}), we have the following inequality:
\begin{equation}
\label{eq:2}
t^{opt}\leq t^*
\end{equation}

Next, we prove that the optimal expected total aggregate return $\Expc(R(t;(u^*(t),{\bl}^{*}(t))))$ for $t{=}t^{opt}$ is same as for $t{=}t^*$, i.e. $\Expc(R(t^{opt};(u^*(t^{opt}),{\bl}^{*}(t^{opt})))){=}m$. We first observe that as $(u^*(t),{\bl}^{*}(t))$ maximizes the expected total aggregate return for a given deadline time $t$, we have the following:
\begin{align}
\label{eq:3}
    \Expc(R(t^{opt};(u^*(t^{opt}),{\bl}^{*}(t^{opt}))))&\geq \Expc(R(t^{opt};(u^{opt},{\bl}^{opt})))\nonumber\\
    &= m
\end{align}
Next, assume that (\ref{eq:3}) holds with strict inequality. Therefore, by (\ref{eq:1}), we have the following:
\begin{align}
\label{eq:5}
\Expc(R(t^{opt};(u^*(t^{opt}),{\bl}^{*}(t^{opt}))))>\Expc(R(t^{*};(u^*(t^*),{\bl}^{*}(t^{*})))).
\end{align}
By the monotonicity of $\Expc(R(t;(u^*(t),{\bl}^{*}(t))))$ with respect to $t$, (\ref{eq:5}) implies $t^{opt}{>}t^*$, which is a contradiction. Hence, $\Expc(R(t^{opt};(u^*(t^{opt}),{\bl}^{*}(t^{opt})))){=}m$. Therefore, using the fact that $t{=}t^*$ is the minimum $t$ such that $\Expc(R(t;(u^*(t),{\bl}^{*}(t)))){=}m$, we have $t^*{\leq} t^{opt}$. Hence, together with (\ref{eq:2}), the claim $t^{*}{=}t^{opt}$ is proved.

\section{Proof of Theorem}
\label{append:theoreProof}
\noindent
Using the computation and communication models presented in (\ref{eq:computeModel}) and (\ref{eq:commModel}) in Section \ref{subsec:delayModel}, we have the following for the execution time for one epoch for node $j{\in}[n{+}1]$:
\begin{align}
	\label{eq:theo1}
	T_j&=T^{(j,1)}_{cmp}+T^{(j,2)}_{cmp}+T^{(j)}_{com-d}+T^{(j)}_{com-u}\nonumber\\
	&=\frac{\wtilde{\ell}_j}{\mu_j}+T^{(j,2)}_{cmp}+\tau_j N^{(j)}_{com},
	\end{align}
	where $N^{(j)}_{com}{\sim}\text{NB}(r{=}2,p{=}1-p_j)$ has negative binomial distribution while $T^{(j,2)}_{cmp}{\sim} \text{E}\left(\frac{\alpha_j\mu_j}{\wtilde{\ell}_j}\right)$ has exponential distribution. Here, we have used the fact that $T^{(j)}_{com-d}$ and $T^{(j)}_{com-u}$ are IID geometric $\text{G}(p)$ random variables and sum of $r$ IID $\text{G}(p)$ is $\text{NB}(r,p)$. Therefore, the probability distribution for $T_j$ is obtained as follows:
	\begin{align}
	\Prob&(T_j\leq t)\nonumber\\
	&=\Prob\left(\frac{\wtilde{\ell}_j}{\mu_j}+T_{cmp}^{(j,2)}+\tau_j N_{com}^{(j)}\leq t\right)\nonumber\\
	&=\sum_{\nu=2}^{\infty}\Prob(N_{com}^{(j)}=\nu)\nonumber\\
	&\quad\quad\quad\quad\cdot\Prob\left(T_{cmp}^{(j,2)}\leq t-\frac{\wtilde{\ell}_j}{\mu_j}-\tau_j N_{com}^{(j)}|N_{com}^{(j)}=\nu\right)\nonumber\\
	&\overset{(a)}=\sum_{\nu=2}^{\infty}\Prob(N_{com}^{(j)}=\nu)\,\Prob\left(T_{cmp}^{(j,2)}\leq t-\frac{\wtilde{\ell}_j}{\mu_j}-\tau_j \nu\right)\nonumber\\
	&\overset{(b)}=\sum_{\nu=2}^{\infty}{U}\left(t-\frac{\wtilde{\ell}_j}{\mu_j} -\tau_j \nu\right)(\nu-1)(1-p_j)^2 p_j^{\nu-2}\nonumber\\
	&\quad\quad\quad\quad\quad\quad\quad\quad\quad\cdot\left(1-e^{{-\frac{\alpha_j\mu_j}{\wtilde{\ell}_j}\left(t-\frac{\wtilde{\ell}_j}{\mu_j}-\tau_j \nu\right)}}\right),
	\end{align}
	where $(a)$ holds due to independence of $T_{cmp}^{(j,2)}$ and $N_{com}^{(j)}$, while in $(b)$, we have used $U(\cdot)$ to denote the unit step function with $U(x){=}1$ for $x{>}0$ and $U(x){=}0$ for $x{\leq}0$. For a fixed $t$, $\Prob(T_j{\leq} t){=}0$ if $t{\leq} 2\tau_j$. For $t{>}2\tau_j$, let $\nu_m{\geq} 2$ satisfy the following criteria:
	\begin{align}
	\label{eq:theo3}
	(t-\tau_j \nu_m)>0,
	(t-\tau_j (\nu_m+1))\leq 0. 
	\end{align}
	Therefore, for $\nu{>}\nu_m$, the terms in $(b)$ are $0$. Finally, as $\Expc({R_{j}(t;\wtilde{\ell}_j)}){=}\wtilde{\ell}_j \Expc({\mathbbm{1}}_{\{T_j\leq t\}}){=}\wtilde{\ell}_j\Prob(T_j{\leq}t)$, we arrive at the result of our Theorem.

{
\section{Proof of Monotonically Increasing Behavior of Optimized Expected Return}
\label{append:monotonicity}
Recall from Section \ref{sec:theory} that for a given deadline time of $t$ at the server, the expectation of the return $R_j(t;\wtilde{\ell}_j){=}\wtilde{\ell}_j {\mathbbm{1}}_{\{T_j\leq t\}}$ for node $j{\in}[n+1]$ satisfies the following: 
\begin{align}
    \Expc&({R_{j}(t;\wtilde{\ell}_j)})\nonumber\\&=\left\{
	\begin{array}{ll}
		\sum_{\nu=2}^{\nu_m}{U}\bigg(t-\frac{\wtilde{\ell}_j}{\mu_j} -\tau_j \nu\bigg)h_{\nu} f_{\nu}(t;\wtilde{\ell}_j)  & \mbox{if } \nu_m{\geq} 2   \\
		0 & \mbox{otherwise } \nonumber
	\end{array}
    \right.
\end{align}
where $U(\cdot)$ is the unit step function with $U(x){=}1$ for $x{>}0$ and $0$ otherwise,\\$f_{\nu}(t;\wtilde{\ell}_j){=}\wtilde{\ell}_j\bigg(1-e^{-\frac{\alpha_j\mu_j}{\wtilde{\ell}_j}(t-\frac{\wtilde{\ell}_j}{\mu_j}-\tau_j \nu)}\bigg)$,\\
$h_{\nu}{=}(\nu-1)(1-p_j)^2p_j^{\nu-2}$, \\
and $\nu_m{\in}\mathbb{Z}$ satisfies $t{-}\tau_j \nu_m{>}0,\, t{-}\tau_j (\nu_m+1){\leq} 0$.

Fix $j{\in}[n+1]$, and consider a fixed load $\wtilde{\ell}_j$ and a given $\nu{\in}\mathbb{Z}$. Then, $f_{\nu}(t;\wtilde{\ell}_j)$ is monotonically increasing in $t$ as $\frac{\partial f_{\nu}(t;\wtilde{\ell}_j)}{\partial t}{=}\alpha_j \mu_j e^{-\frac{\alpha_j\mu_j}{\wtilde{\ell}_j}(t-\frac{\wtilde{\ell}_j}{\mu_j}-\tau_j \nu)}{>}0$ for all $t{>}0$. Furthermore, by definition, $\nu_m$ is monotonically increasing in deadline time $t$. Therefore, total number of terms in the expression for $\Expc({R_{j}(t;\wtilde{\ell}_j)})$ increases monotonically with $t$, and each of those terms increases monotonically with $t$. Thus, for a fixed $\wtilde{\ell}_j$, $\Expc({R_{j}(t;\wtilde{\ell}_j)})$ is monotonically increasing in $t$.

Consider two different deadline times $t{=}t_1$ and $t{=}t_2$ with $t_2{>}t_1$. Based on the discussion in Section \ref{sec:theory}, let $\wtilde{\ell}_j{=}\ell_j^*(t_1)$ be the optimal load that maximizes $\Expc({R_{j}(t_1;\wtilde{\ell}_j)})$ and let $\Expc({R_{j}(t_1;\ell_j^*(t_1))})$ be the corresponding optimized expected return. Similarly, let $\wtilde{\ell}_j{=}\ell_j^*(t_2)$ be the optimal load that maximizes $\Expc({R_{j}(t_2;\wtilde{\ell}_j)})$ and let $\Expc({R_{j}(t_2;\ell_j^*(t_2))})$ be the corresponding optimized expected return. Since $t_2{>}t_1$ and expected return is monotonically increasing with $t$, we have  $\Expc({R_{j}(t_2;\ell_j^*(t_1))}){\geq} \Expc({R_{j}(t_1;\ell_j^*(t_1))})$. Since $\Expc({R_{j}(t_2;\ell_j^*(t_2))})$ is the optimal expected return for $t_2$, it follows that $\Expc({R_{j}(t_2;\ell_j^*(t_2))}){\geq}\Expc({R_{j}(t_2;\ell_j^*(t_1))}){\geq} \Expc({R_{j}(t_1;\ell_j^*(t_1))})$. Therefore, the optimized expected return $\Expc({R_{j}(t;\ell_j^*(t))})$ is monotonically increasing in the deadline time $t$.
}
\section{One-shot Solution for AWGN}
\label{append:oneshot}
For node $j{\in}[n{+}1]$, consider the function  $f_{\nu}(t;\wtilde{\ell}_j){=}\wtilde{\ell}_j(1-e^{-\frac{\alpha_j\mu_j}{\wtilde{\ell}_j}(t-\frac{\wtilde{\ell}_j}{\mu_j}-\tau_j \nu)})$ for $\nu{\in}\{2,\ldots,\nu_m\}$. We recall that $f_{\nu}(t;\wtilde{\ell}_j)$ is strictly concave for $\wtilde{\ell}_j{>}0$. Furthermore, $f_{\nu}(t;\wtilde{\ell}_j){\leq}0$ for $\wtilde{\ell}_j{\geq}{\mu_j(t-\tau_j \nu)}$. Solving for $f'_{\nu}(t;\wtilde{\ell}_j){=}0$, we obtain the optimal load maximizing $f_{\nu}(t;\wtilde{\ell}_j)$ as follows:  
\begin{align}
\label{eq:optLfn}
    \ell_{j}^*(t,\nu) = -\frac{\alpha_j \mu_j}{W_{-1}(-e^{-(1+\alpha_j)})+1}(t-\nu\tau_j),
\end{align}
where $W_{-1}(\cdot)$ is the minor branch of Lambert $W$-function, where the Lambert $W$-function is the inverse function of $f(W){=}We^W$. 

Next, consider the special case of AWGN channel in Section \ref{sec:theory}, for which the expected return for node $j{\in}[n{+}1]$ simplifies as follows: 
\begin{equation}
\Expc({R_{j}(t;\wtilde{\ell}_j)})={{U}\left(t-\frac{\wtilde{\ell}_j}{\mu_j} -2\tau_j\right)}f_{2}(t;\wtilde{\ell}_j).
\end{equation} 
When $t{\leq} 2\tau_j$, $\Expc({R_{j}(t;\wtilde{\ell}_j)}){=}0$, thus $\ell_j^*(t){=}0$. For $t{>}2\tau_j$, using (\ref{eq:optLfn}) and the fact that $\wtilde{\ell}_j$ is upper bounded by $\ell_j$, $\ell_j^*(t){=}\min\{\ell_j^*(t,2),\ell_j\}$, where $\ell_j^*(t,2)$ is as follows: 
\begin{align}
\label{eq:optLf2}
\ell_j^*(t,2)&=s_j(t-2\tau_j),
\end{align}
where $s_j{=}-\frac{\alpha_j \mu_j}{W_{-1}\left(-e^{-(1+\alpha_j)}\right)+1}$. As $\ell_j^*(t,2)$ is strictly increasing, there is a unique deadline time for which $\ell_j^*(t,2){=}\ell_j$. Let $t{=}\zeta_j$ be the deadline time for which this holds. Then, using (\ref{eq:optLf2}), we have the following for  $\zeta_j$:
\begin{equation}
    \zeta_j = \frac{\ell_j}{s_j}+2\tau_j.
\end{equation}
Thus, we have the following closed form expression for $\ell_j^*(t)$:
\begin{align}
\label{eq:1closedFormClientLoad1}
    \ell_j^*(t)=\left\{
	\begin{array}{ll}
		0  & \mbox{if } t\leq 2\tau_j   \\
		s_j(t-2\tau_j) & \mbox{if } 2\tau_j<t\leq \zeta_j \\
		\ell_j & \mbox{otherwise }
	\end{array}
    \right.
\end{align}
Furthermore, by (\ref{eq:optLf2}), the optimal expected return from $j$-th node for deadline time $2\tau_j{<}t{\leq} \zeta_j$ can be simplified as follows:
\begin{align}
\Expc(R_j(t;\ell_j^*(t)))&=\ell_j^*(t)\left(1-e^{-\frac{\alpha_j\mu_j}{\ell_j^*(t)}\left(t-\frac{\ell_j^*(t)}{\mu_j}-2\tau_j\right)}\right),\nonumber\\
&=s_j(t-2\tau_j)\left(1-e^{-\alpha_j\left(\frac{\mu_j}{s_j}-1\right)}\right),\nonumber\\
&=\wtilde{s}_j(t-2\tau_j),
\end{align}
where $\wtilde{s}_j{=}s_j \left(1-e^{-\alpha_j\left(\frac{\mu_j}{s_j}-1\right)}\right)$. Thus, we have the following closed form expression for the optimal expected return:
\begin{align}
\label{eq:closedFormClientRet1}
    \Expc&(R_j(t;\ell_j^*(t)))\nonumber\\
    &=\left\{
	\begin{array}{ll}
		0  & \mbox{if } t\leq 2\tau_j   \\
		\wtilde{s}_j(t-2\tau_j) & \mbox{if } 2\tau_j<t\leq \zeta_j \\
		\ell_j\left(1-e^{-\frac{\alpha_j\mu_j}{\ell_j}\left(t-\frac{\ell_j}{\mu_j}-2\tau_j\right)}\right) & \mbox{otherwise }
	\end{array}
    \right.
\end{align}
Thus, we have a closed form for the maximum expected total aggregate return from the nodes till deadline time $t$ as follows:
\begin{align}
\label{eq:totalExpRet1}
\Expc(R(t;\bl^*(t)))&=\sum_{\substack{j\in[n+1]\\
                  2\tau_j<t\leq \zeta_j}} \wtilde{s}_j(t-2\tau_j)
        \nonumber\\
        &+\sum_{\substack{j\in[n+1]\\
                  \zeta_j<t}}\ell_j\left(1-e^{-\frac{\alpha_j\mu_j}{\ell_j}\left(t-\frac{\ell_j}{\mu_j}-2\tau_j\right)}\right),
\end{align}
which is monotonically increasing in $t$.

{
\section{Towards Convergence Analysis of CodedFedL}
\label{append:proofConvergence}
For proving convergence of CodedFedL, we consider $u^*(t^*)$ to be large, and make the following assumption for simplification: 
\begin{equation}
    \frac{\bG^T \bG}{u^*(t^*)} = \mathbf{I}_m.
\end{equation}
The key motivation for our assumption is the observation that by weak law of large numbers, in the limit that the coding redundancy $u^*(t^*){\rightarrow} \infty$, each diagonal entry in $\frac{\bG^T \bG}{u^*(t^*)}$ converges to $1$ in probability, while each non-diagonal entry converges to $0$ in probability. Furthermore, as we demonstrate via numerical experiments in Section \ref{sec:experiments}, the convergence curve as a function of iteration for CodedFedL significantly overlaps that of the naive uncoded where the server waits to aggregate the results of all the clients. Hence, for simplifying our analysis, we assume in the remaining proof that $\frac{\bG^T \bG}{u^*(t^*)} = \mathbf{I}_m$. The general analysis will be addressed in future work. 

In the following, we list the remaining assumptions in our analysis:
\begin{itemize}
\item Assumption 1: Model parameter space $\mathcal{W}{\subseteq} \mathbb{R}^{q\times c}$ is a closed and convex set.
\item Assumption 2: $\sup_{\btheta{\in}\mathcal{W}}\norm{\btheta-\btheta^{(0)}}_F^2{\leq} R^2$, where $\btheta^{(0)}{\in} \mathcal{W}$ is given.  
\item Assumption 3: $\norm{\frac{1}{\ell_j^*(t^*)}\wtilde{\bX}^{(j)T}(\wtilde{\bX}^{(j)}\btheta-\wtilde{\bY}^{(j)})}_F^2{\leq} B_j$ for all $\btheta{\in}\mathcal{W}$ for client $j{\in}[n]$.
\item Assumption 4: $\text{Max}\{\text{singular values of }\what{\bX}^{(j)}\}{\leq} L_j$, for client $j{\in}[n]$.
\item Assumption 5: $\Prob(T_C\leq t^*){=}1$, i.e. the gradient over the coded data is available at the MEC server by $t^*$ with probability $1$.
\end{itemize}
Under the above assumptions, for a given model $\btheta{\in}\mathcal{W}$, the stochastic gradient obtained by the MEC server by the deadline time $t^*$ is as follows:
\begin{align}
    &\bg_M(\btheta)\nonumber\\
    &=\frac{1}{m}(\bg_C(\btheta)+\bg_U(\btheta)),\nonumber\\
    &=\frac{1}{m}\Big(\what{\bf X}^T {\bf W}^T{\bf W}(\what{\bX}{\btheta}-\bY)\nonumber\\
    &\quad\quad\quad\quad\quad\quad+\sum_{j=1}^n{\mathbbm{1}}_{\{T_j\leq t^{*}\}} \wtilde{\bX}^{(j)T}(\wtilde{\bX}^{(j)}\btheta-\wtilde{\bY}^{(j)})\Big),\nonumber\\
    &= \frac{1}{m}\Big(\sum_{j=1}^{n}\sum_{\substack{k\in[\ell_j]\\(\what{\bx}^{(j)}_k,{\by}^{(j)}_k)\in\wtilde{D}_{j}}}(1-\Prob(T_j{\leq}t^*))\nonumber\\
    &\quad\quad\quad\quad\quad\quad\quad\quad\quad\quad\quad\quad\cdot\what{\bf x}^{(j)T}_k(\what{\bf x}^{(j)}_k {\btheta} - {\by}^{(j)}_k)\nonumber\\&                          \quad\quad+\sum_{j=1}^{n}\sum_{\substack{k\in[\ell_j]\\(\what{\bx}^{(j)}_k,{\by}^{(j)}_k)\in\what{D}_j{\setminus}\wtilde{D}_j}}\what{\bf x}^{(j)T}_k(\what{\bf x}^{(j)}_k {\btheta} - {\by}^{(j)}_k)\nonumber\\&+    \sum_{j=1}^{n}\sum_{\substack{k\in[\ell_j]\\(\what{\bx}^{(j)}_k,{\by}^{(j)}_k)\in\wtilde{D}_{j}}}{\mathbbm{1}}_{\{T_j\leq t^{*}\}}\what{\bf x}^{(j)T}_k(\what{\bf x}^{(j)}_k {\btheta} - {\by}^{(j)}_k) \Big).
\end{align}
Averaging over the stochastic conditions of compute and communication, we have the following for a given $\btheta{\in}\mathcal{W}$:
\begin{align}
&\Expc(\bg_M(\btheta))\nonumber\\ 
&=\frac{1}{m}\Big(\sum_{j=1}^{n}\sum_{\substack{k\in[\ell_j]\\(\what{\bx}^{(j)}_k,{\by}^{(j)}_k)\in\wtilde{D}_{j}}}(1-\Prob(T_j{\leq}t^*))\what{\bf x}^{(j)T}_k(\what{\bf x}^{(j)}_k {\btheta} - {\by}^{(j)}_k)\nonumber\\&                          \quad+\sum_{j=1}^{n}\sum_{\substack{k\in[\ell_j]\\(\what{\bx}^{(j)}_k,{\by}^{(j)}_k)\in\what{D}_j{\setminus}\wtilde{D}_j}}\what{\bf x}^{(j)T}_k(\what{\bf x}^{(j)}_k {\btheta} - {\by}^{(j)}_k)\nonumber\\
&\quad+\sum_{j=1}^{n}\sum_{\substack{k\in[\ell_j]\\(\what{\bx}^{(j)}_k,{\by}^{(j)}_k)\in\wtilde{D}_{j}}} \Prob(T_j\leq t^{*}) \what{\bf x}^{(j)T}_k(\what{\bf x}^{(j)}_k {\btheta} - {\by}^{(j)}_k) \Big),\nonumber\\
    &= \frac{1}{m}\sum_{j=1}^n\sum_{k=1}^{\ell_j}\what{\bf x}^{(j)T}_k(\what{\bf x}^{(j)}_k {\btheta} - {\by}^{(j)}_k),\nonumber\\
    &= {\nabla_{\btheta}} \Big(\frac{1}{2m}\sum_{j=1}^n\norm{\what{\bX}^{(j)}\btheta-\bY^{(j)}}_F^2\Big),\nonumber\\
    &= \bg(\btheta)
\end{align}
Thus, the variance of $\bg_M(\btheta)$ for a given $\btheta{\in}\mathcal{W}$ can be bounded as follows:
\begin{align}
\Expc&(\norm{\bg_M(\btheta)-\Expc(\bg_M(\btheta))}_F^2)\\
    &=\Expc\Big(\Big|\!\Big|   \frac{1}{m}\sum_{j=1}^n    ({\mathbbm{1}}_{\{T_j\leq t^{*}\}}-\Prob(T_j{\leq}t^*))
    \nonumber\\
    &\quad\quad\quad\quad\quad\quad\cdot\sum_{\substack{k\in[\ell_j]\\(\what{\bx}^{(j)}_k,{\by}^{(j)}_k)\in\wtilde{D}_{j}}}                    \what{\bf x}^{(j)T}_k(\what{\bf x}^{(j)}_k {\btheta} - {\by}^{(j)}_k)\Big|\!\Big|^2_F\Big),\nonumber\\
    &\overset{(a)}=\frac{1}{m^2}\sum_{j=1}^n \Prob(T_j{\leq}t^*)(1-\Prob(T_j{\leq}t^*)) \nonumber\\
    &\quad\quad\quad\quad\quad\quad\cdot\Big|\!\Big|\sum_{\substack{k\in[\ell_j]\\(\what{\bx}^{(j)}_k,{\by}^{(j)}_k)\in\wtilde{D}_{j}}}                    \what{\bf x}^{(j)T}_k(\what{\bf x}^{(j)}_k {\btheta} - {\by}^{(j)}_k)\Big|\!\Big|^2_F,\nonumber\\
    &=\frac{1}{m^2}\sum_{j=1}^n \Prob(T_j{\leq}t^*)(1-\Prob(T_j{\leq}t^*))\\
    &\quad\quad\quad\quad\quad\quad\cdot\norm{\wtilde{\bX}^{(j)T}(\wtilde{\bX}^{(j)}\btheta-\wtilde{\bY}^{(j)})}_F^2\nonumber\\
    &\leq \sum_{j=1}^n \frac{({\ell}^*_j(t^*))^2}{m^2}
    \big\lVert\frac{1}{{\ell}_j^*(t^*)}\wtilde{\bX}^{(j)T}(\wtilde{\bX}^{(j)}\btheta-\wtilde{\bY}^{(j)})\big\rVert_F^2\\ 
    &\leq\sum_{j=1}^n B_j=B,
\end{align}
where in $(a)$, we have used independence of the events ${\mathbbm{1}}_{\{T_{j_1}\leq t^{*}\}}$ and ${\mathbbm{1}}_{\{T_{j_2}\leq t^{*}\}}$ for distinct clients $j_1{\in}[n]$ and $j_2{\in}[n]$. Furthermore, for $\btheta_1{\in}\mathcal{W}$ and $\btheta_2{\in}\mathcal{W}$, we have the following bound for smoothness:\\
\begin{align}
    \lVert \bg(\btheta_1)&-\bg(\btheta_2)\rVert_F \nonumber\\
    &=\Big|\!\Big| \frac{1}{m}\sum_{j=1}^n\Big(\what{\bX}^{(j)T}(\what{\bX}^{(j)}\btheta_1-\what{\bY}^{(j)})\nonumber\\
    &\quad\quad\quad\quad\quad\quad-\what{\bX}^{(j)T}(\what{\bX}^{(j)}\btheta_2-\what{\bY}^{(j)})\Big)\Big|\!\Big|_F,\nonumber\\
    &=\frac{1}{m}\Big\lVert{\sum_{j=1}^n \Big(\what{\bX}^{(j)T}\what{\bX}^{(j)}(\btheta_1- \btheta_2)  \Big)}\Big\rVert_F,\nonumber\\
    &\leq \frac{1}{m}\sum_{j=1}^n\Big\lVert{ \Big(\what{\bX}^{(j)T}\what{\bX}^{(j)}(\btheta_1- \btheta_2)  \Big)}\Big\rVert_F,\nonumber\\
    &\overset{a}\leq \frac{1}{m}\sum_{j=1}^n\big\lVert \what{\bX}^{(j)T}\what{\bX}^{(j)}\big\rVert_2 \big \lVert\btheta_1- \btheta_2\big\rVert_F,\nonumber\\
    &\leq\frac{1}{m}\sum_{j=1}^n L_j^2\norm{\btheta_1-\btheta_2}_F=L\norm{\btheta_1-\btheta_2}_F.
\end{align}
In $(a)$, $\norm{A}_2$ denotes the spectral norm of matrix $A$. Therefore, by Theorem $2.1$ in \cite{qsgd}, for a total number of $r^{max}$ iterations and a constant learning rate of $\mu^{(r)}{=}\frac{1}{L+1/\gamma}$ with $\gamma{=}\sqrt{\frac{2 R^2}{Br^{max}}}$, we have the following result:
\begin{align}
    &\Expc\Big(\frac{1}{2m}\sum_{j=1}^n\norm{\what{\bX}^{(j)}\btheta^{1:r^{max}}-\bY^{(j)}}_F^2\Big)\nonumber\\
    &\quad\quad\quad\quad-\min_{\btheta\in\mathcal{W}}\frac{1}{2m}\sum_{j=1}^n\norm{\what{\bX}^{(j)}\btheta-\bY^{(j)}}_F^2\nonumber\\
    &\quad\quad\quad\quad\quad\quad\quad\quad\leq R\sqrt{\frac{2 B}{r^{max}}}+\frac{L R^2}{r^{max}},
\end{align}
where $\btheta^{1:r^{max}}{=}\frac{1}{r^{max}}\sum_{r=1}^{r^{max}} \btheta^{(r)}$. Hence, for achieving an error less than a given $\epsilon{>}0$, the iteration complexity of CodedFedL is $r^{max}=\mathcal{O}(R^2 \max(\frac{2B}{\epsilon^2},\frac{L}{\epsilon}))$.
}

{
\section{Privacy Budget for CodedFedL}
\label{app:privacy}
We utilize $\epsilon$-mutual-information differential privacy (MI-DP)  metric, as proposed in \cite{cuff2016differential}, for finding privacy leakage in CodedFedL. For completeness, we first provide the definition of $\epsilon$-MI-DP {(shown to be stronger than the standard $(\epsilon,\delta)$-DP metric)} as presented in \cite{showkatbakhsh2018privacy}.
\begin{itemize}
\item $\epsilon$-Mutual-Information Differential Privacy: Let $D^N=(D_1,\ldots,D_N)$ be a database with $N$ entries. $D^N$ returns a query as per a randomized mechanism $Q(\cdot)$. Let $D^{-i}$ be the database with all entries except $D_i$. Then, the randomized mechanism satisfies $\epsilon$-MI-DP if the following is satisfied:
\begin{equation}
    \sup_{i,\Prob(\mathbf{D}^N)} I(\mathbf{D}_i;Q(\mathbf{D}^N)|\mathbf{D}^{-i})\leq\epsilon \text{ bits},
\end{equation}
where the supremum is taken over all distributions on $\mathbf{D}^N$.
\end{itemize}
Next, leveraging the result for random linear projections in \cite{showkatbakhsh2018privacy}, we can calculate the privacy budget required for sharing the local parity dataset $(\widebreve{\bX}^{(j)},\widebreve{\bY}^{(j)})$ for a given client $j{\in}[n]$. As we aim to preserve the privacy of each entry of $\what{\bX}^{(j)}$, we need to compute the required privacy budget with respect to the largest diagonal entry of the scaling matrix $\bW_j$. Therefore, replacing $\bW_j$  by an identity matrix, we equivalently consider the privacy leakage for sharing   $(\widebreve{\bX}^{(j)}{=} \bG_j \what{\bX}^{(j)},\,
\widebreve{\bY}^{(j)}{=} \bG_j {\bY}^{(j)})$ (see Sections \ref{encode} and \ref{sec:weight} for details). Furthermore, we assume that the entries of $\bG_j$ are drawn independently from a standard normal distribution. Then, based on the result for $\epsilon$-MI-DP from  Section III-B of \cite{showkatbakhsh2018privacy}, CodedFedL needs to allocate $\epsilon_j$ privacy budget for sharing $u^*(t^*)$ number of local parity data $(\widebreve{\bX}^{(j)},\widebreve{\bY}^{(j)})$ to the MEC server, where $\epsilon_j$ is given by:
\begin{align}
\epsilon_j = \frac{1}{2}\log_2\Big(1+\frac{u^*(t^*)}{f^2(\what{\bX}^{(j)})}\Big),  
\end{align}
where $$f(\what{\bX}^{(j)})=\min_{k_2\in[q]}\sqrt{\sum_{k_1=1}^{\ell_j}|\what{\bx}_{k_1}^{(j)}(k_2)|^2-\max_{k_3\in[\ell_j]}|\what{\bx}_{k_3}^{(j)}(k_2)|^2}.$$ Here, we have used $\what{\bx}_{i}^{(j)}(k)$ to denote the value of $i$-th data point corresponding to the $k$-th feature in raw database $\what{\bX}^{(j)}$. Intuitively, when the raw data distribution is concentrated along a small number of features, the value of $f(\what{\bX}^{(j)})$ is small and a larger privacy budget is required for generating coded data to effectively hide those vulnerable features. In contrast, when raw data distribution is uniform in feature space, very little information is leaked by the parity data generated in CodedFedL.

}

\begin{IEEEbiography}
    [{\includegraphics[width=1in,height=1.25in,clip,keepaspectratio]{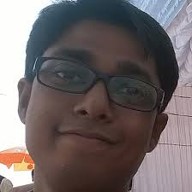}}]{Saurav Prakash} (IEEE Graduate Student Member)
received the Bachelor of Technology degree in Electrical
Engineering from the Indian Institute of Technology (IIT), Kanpur, India, in 2016. He is currently pursuing the Ph.D. degree in Electrical and Computer Engineering with the University of Southern California (USC), Los Angeles. He was a finalist in the Qualcomm Innovation Fellowship program in 2019. His research interests include information theory and data analytics with applications in large-scale machine learning and edge computing. He received the USC Annenberg Graduate Fellowship in 2016 and was one of the Viterbi-India fellows in Summer 2015.
\end{IEEEbiography}

\begin{IEEEbiography}[{\includegraphics[width=1in,height=1.33in,clip,keepaspectratio]{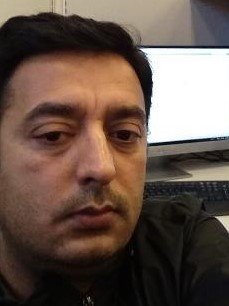}}]{Sagar Dhakal} (IEEE Senior Member) completed high school in Kathmandu, Nepal. He received the Bachelor degree in Electrical and Electronics Engineering from Birla Institute of Technology, India, in 2001, and the MS and the PhD degrees in Electrical Engineering from University of New Mexico, USA in 2003 and 2006, respectively. He is currently working as a Principal Engineer at Broadcom Corporation in San Jose, California. Previously, Sagar has worked at Intel Labs, RIM, Nortel, US Naval Research Lab, and a cloud-RAN start-up. His current research interests are in the field of signal processing, wireless communication and machine learning.
\end{IEEEbiography}

\begin{IEEEbiography}[{\includegraphics[width=1in,height=1.25in,clip,keepaspectratio]{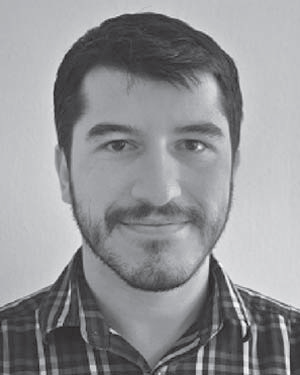}}]{Mustafa Riza Akdeniz} (S'09, M'16) received the B.S. degree in electrical and electronics engineering from Bogazici University, Istanbul, Turkey, in 2010 and the Ph.D. degree in electrical and computer engineering at New York University Tandon School of Engineering, Brooklyn, NY in 2016. He is working as a research scientist for Intel Labs in Santa Clara, CA. His research interests include wireless channel modeling, information theory, distributed learning.
\end{IEEEbiography}

\begin{IEEEbiography}[{\includegraphics[width=1in,height=1.25in,clip,keepaspectratio]{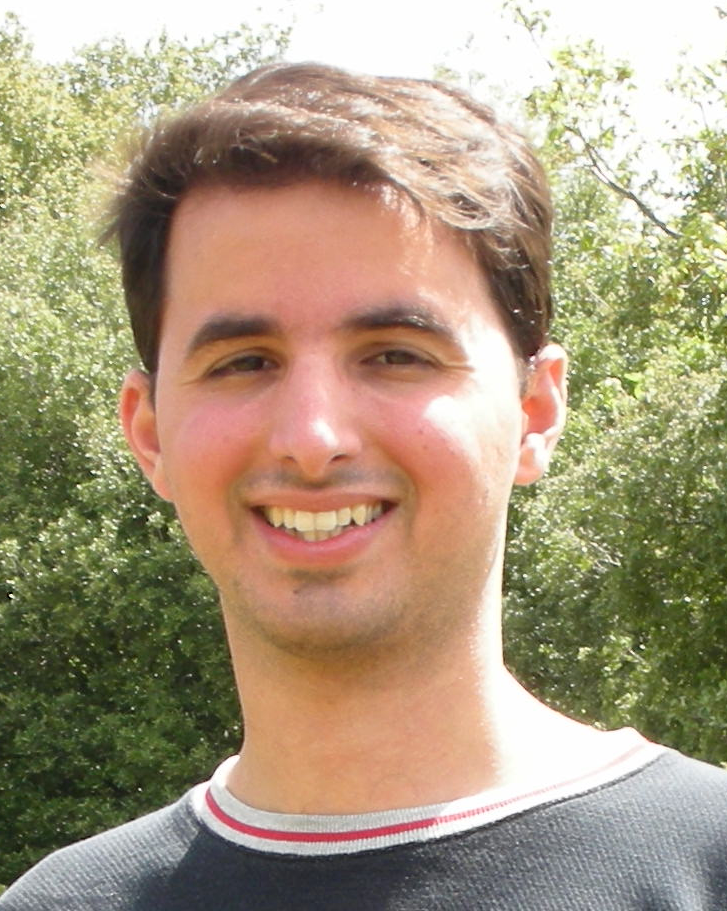}}]{Yair Yona} received a B.Sc. in Electrical Engineering (magna cum laude), an M.Sc. in Electrical Engineering (magna cum laude), and a Ph.D. in Electrical Engineering, all from Tel-Aviv University, Israel, in 2005, 2009 and 2014, respectively. He is currently a staff engineer at Qualcomm, San Jose, CA.  

Between 2017 and 2019 he was a research scientist at Intel Labs. From 2015 to 2017 he was a postdoctoral scholar in the Department of Electrical Engineering at the University of California Los Angeles, Los Angeles, CA. 

Dr. Yona was a recipient of the Intel award for excellence in academic studies and research (2009), a Motorola Scholarship in the field of advanced communication (2009), and the Weinstein Prize (2010, 2014) for research in the area of signal processing.
\end{IEEEbiography}

\begin{IEEEbiography}[{\includegraphics[width=1.0in,height=1.0in,clip,keepaspectratio]{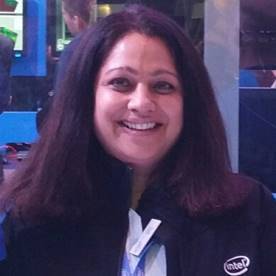}}]{Shilpa Talwar} (Intel Fellow) is director of wireless multi-communication systems in the Intel Labs organization at Intel Corporation. She leads a research team in the Wireless Communications Laboratory focused on advancements in ultra-dense multi-radio network architectures and associated technology innovations. Her research interests include multi-radio convergence, interference management, mmWave beamforming, and applications of machine learning and artificial intelligence (AI) techniques to wireless networks. While at Intel, she has contributed to IEEE and 3GPP standard bodies, including 802.16m, LTE-advanced, and 5G NR.  She is co-editor of book on 5G “Towards 5G: Applications, requirements and candidate technologies.” Prior to Intel, Shilpa held several senior technical positions in wireless industry working on a wide-range of projects, including algorithm design for 3G/4G and WLAN chips, satellite communications, GPS, and others. Shilpa graduated from Stanford University in 1996 with a Ph.D. in Applied mathematics and an M.S. in Electrical Engineering. She is the author of 70 technical publications and holds 60 patents.
\end{IEEEbiography}

\begin{IEEEbiography}[{\includegraphics[width=1.0in,height=1.33in,clip,keepaspectratio]{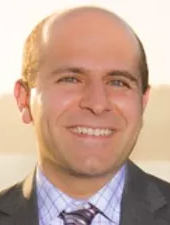}}]{Salman Avestimehr} (IEEE Fellow)
received the B.S. degree in Electrical Engineering from the Sharif University of Technology in 2003, and the M.S. and Ph.D. degrees in Electrical Engineering and Computer Science from the University of California, Berkeley, in 2005 and 2008, respectively. He is currently a Professor and the Director of the Information Theory and Machine Learning (vITAL) Research Laboratory, Electrical and Computer Engineering Department, University of Southern California. His research interests include information theory, coding theory, and large-scale distributed computing and machine learning. 

Dr. Avestimehr has received a number of awards for his research, including the James L. Massey Research and Teaching Award from the IEEE Information Theory Society, the Information Theory Society and Communication Society Joint Paper Award, the Presidential Early Career Award for Scientists and Engineers (PECASE) from the White House, the Young Investigator Program (YIP) Award from the U.S. Air Force Office of Scientific Research, the National Science Foundation CAREER Award, the David J. Sakrison Memorial Prize, and several best paper awards at Conferences. He has been an Associate Editor of the IEEE TRANSACTIONS ON INFORMATION THEORY. He was also the General Co-Chair of the 2020 International Symposium on Information Theory (ISIT).
\end{IEEEbiography}

\begin{IEEEbiography}[{\includegraphics[width=1.0in,height=1.1in,clip,keepaspectratio]{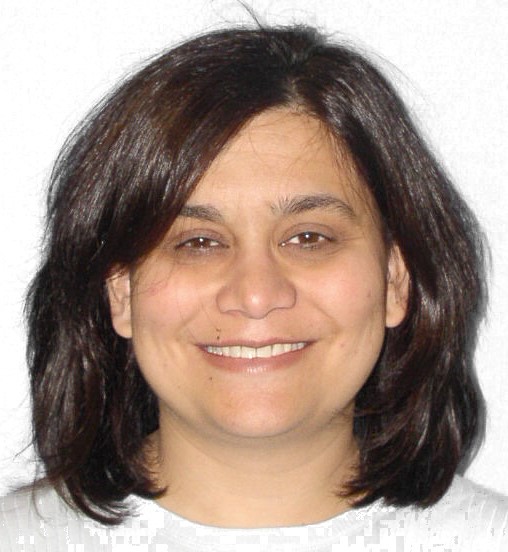}}]{Nageen Himayat}
is a Director and Principal Engineer with Intel Labs, where she conducts research on distributed learning and data centric protocols over 5G/5G+wireless networks.   Her research contributions span areas such as machine learning for wireless, millimeter wave and multi-radio heterogeneous networks, cross layer radio resource management, and non-linear signal processing techniques. She has authored over 300 technical publications, contributing to several IEEE peer-reviewed publications, 3GPP/IEEE standards, as well as numerous patent filings.

Prior to Intel, Dr. Himayat was with Lucent Technologies and General Instrument Corp, where she developed standards and systems for both wireless and wire-line broadband access networks. Dr. Himayat obtained her B.S.E.E degree from Rice University, and her Ph.D. degree from the University of Pennsylvania.  She also holds an MBA degree from the Haas School of Business at University of California, Berkeley.  

\end{IEEEbiography}

\end{document}